\documentclass[traditabstract]{aa}

\usepackage{amsmath}
\usepackage{graphicx}
\usepackage{txfonts}
\usepackage{natbib}

\begin{document}

\title{Dusty shells surrounding the carbon variables S~Scuti and RT~Capricorni}

\author{M.~Me\v{c}ina\inst{\ref{inst1}} \and F. Kerschbaum\inst{\ref{inst1}} \and M.A.T. Groenewegen\inst{\ref{inst2}} \and R. Ottensamer\inst{\ref{inst1}} \and J.A.D.L. Blommaert\inst{\ref{inst3},\ref{inst4}} \and \\A. Mayer\inst{\ref{inst1}} \and L. Decin\inst{\ref{inst3}} \and A. Luntzer\inst{\ref{inst1}} \and B. Vandenbussche\inst{\ref{inst3}} \and Th. Posch\inst{\ref{inst1}} \and C. Waelkens\inst{\ref{inst3}}}

\institute{Department of Astrophysics, University of Vienna, T\"{u}rkenschanzstra{\ss}e 17, 1180 Vienna, Austria\\\email{marko.mecina@univie.ac.at}\label{inst1} \and Koninklijke Sterrenwacht van Belgi\"{e}, Ringlaan 3, 1180 Brussels, Belgium\label{inst2} \and Instituut voor Sterrenkunde, KU Leuven, Celestijnenlaan, 200D, 3001 Leuven, Belgium\label{inst3} \and Department of Physics and Astrophysics, Vrije Universiteit Brussel, Pleinlaan 2, 1050 Brussels, Belgium\label{inst4}}

\date{Submitted 2013}

\abstract
{For the \emph{Mass-loss of Evolved StarS} (MESS) programme, the unprecedented spatial resolution of the PACS photometer on board the \emph{Herschel} space observatory was employed to map the dusty environments of asymptotic giant branch (AGB) and red supergiant (RSG) stars. Among the morphologically heterogeneous sample, a small fraction of targets is enclosed by spherically symmetric detached envelopes. Based on observations in the 70\,$\mu$m and 160\,$\mu$m wavelength bands, we investigated the surroundings of the two carbon semiregular variables S~Sct and RT~Cap, which both show evidence for a history of highly variable mass-loss. S~Sct exhibits a bright, spherically symmetric detached shell, 138$\arcsec$ in diameter and co-spatial with an already known CO structure. Moreover, weak emission is detected at the outskirts, where the morphology seems indicative of a mild shaping by interaction
of the wind with the interstellar medium, which is also supported by the stellar space motion. Two shells are found around RT~Cap that were not known so far in either dust emission or from molecular line observations. The inner shell with a diameter of 188$\arcsec$ shows an almost immaculate spherical symmetry, while the outer $\sim5\arcmin$ structure is more irregularly shaped. \emph{MoD}, a modification of the DUSTY radiative transfer code, was used to model the detached shells. Dust temperatures, shell dust masses, and mass-loss rates are derived for both targets.}

 \keywords{Stars: AGB and post-AGB -- stars: carbon -- stars: mass-loss -- infrared: stars -- stars: individual: S~Sct -- stars: individual: RT~Cap}
 
 \maketitle

\section{Introduction}
Stars of low and intermediate mass ($\sim$0.8--8$\,\mathrm{M_{\sun}}$) lose considerable amounts of matter through a slow stellar wind while they are on the asymptotic giant branch (AGB). A consequence of the sometimes heavy mass-loss ($\sim$10$^{-8}$ to 10$^{-4}$~$\mathrm{M_{\sun}}$~yr$^{-1}$), common to
C-type (C/O$>$1) AGB stars \citep{Mattsson2007}, is the formation of circumstellar envelopes (CSEs), which consist of gas and dust particles. Understanding the involved processes is crucial, considering the large contribution of AGB stars to the interstellar medium,
with which they essentially influence the chemical evolution of the galaxy \citep[e.g.][]{Schroder2001}. The overall picture of the mass-loss phenomenon on the AGB is quite well established \citep[e.g.][]{Habing1996}, particularly for C-stars. Some details, however, such as the underlying mechanisms and their temporal evolution, are not yet completely understood, together with their dependence on stellar parameters \citep[e.g.][and references therein]{Mattsson2010}.

The ejected dust in the CSEs obscures the central source and dramatically affects the radiation from the star. Based on photometric data from the IRAS space telescope, \citet{Veen1988} were able to identify several AGB stars through their 60\,$\mu$m excess. Some targets, classified in their group VIa, are surrounded by very cold dust at large distances from the central star, so-called detached shells. 
Highly resolved images -- and thus morphological studies -- of the circumstellar environments, however, were not achievable at far-IR (FIR) wavelengths (that picture the cold dust component) because of the limited size of the IRAS telescope aperture (57\,cm). Still, \citet{Young1993} examined hundreds of IRAS sources and found evidence of extended structures (mostly not detached shells) around 78 of them. The situation regarding spatial resolution did not change dramatically with the following ISO mission, although the improved sensitivity led to additional detections. \citet{Izumiura1996} found a detached dust shell around the carbon star \object{Y~CVn,
for example}. More recently, Spitzer and AKARI provided better insight into the dusty circumstellar environment and its possible interaction with the interstellar medium (ISM) \citep[e.g.][]{Ueta2006,Ueta2008,Izumiura2009,Geise2010}. The latest addition to the observational facilities for the far-infrared is the \textit{Herschel Space Observatory} \citep{Pilbratt2010}. An early demonstration of its capabilities was given by \citet{Kerschbaum2010}, who presented a spatially highly resolved view of the detached large-scale dust morphology of three carbon stars. These objects were only a small part of a more comprehensive observing programme \citep[MESS, see][]{Groenewegen2011}, investigating the more extended circumstellar dust structures of evolved stars. A first, a mainly morphological analysis of all observed targets was conducted by \citet{Cox2012}, which showed a rich diversity in shapes, among them previously undetected detached shells. So far, 12 detached-shell objects are known in total, 10 of which have been detected by Herschel.

The research on the gas component of the CSEs is mainly conducted by millimetre and sub-mm radio observations. Primarily, emission lines of CO are studied. The high bond energy of the molecule prevents its quick (photo-)dissociation and thus makes it traceable over a large distance range from the central star. While signatures identified in the spectral profiles of single-dish surveys revealed detached shells around a number of objects \citep{Olofsson1996}, interferometric measurements provided detailed maps of their structure \citep{Lindqvist1999,Olofsson2000}. Very recently, a detailed map using ALMA revealed a remarkable spiral structure around \object{R Scl} due to a stellar companion \citep{Maercker2012}, where previously only a detached shell was found. In addition
to CO, single-dish spectra and interferometric maps of other molecules such as HCN, CN, and CS were obtained as well \citep{Olofsson1993,Lindqvist1996,Wong2004}.

\citet{Olofsson2000} suggested that the geometrically thin shells are a result of strong modulations of the mass-loss rate on a short time-scale. Additionally, the structures could be enhanced by a young, fast wind interacting with the remnants of a slower previous expulsion \citep{Schoier2005}. According to \citet{Wareing2006}, detached shells could also be explained by a wind-ISM interaction for some objects.


In this paper we present FIR-imaging results from observations of the two carbon stars \object{S~Sct} and \object{RT~Cap}, taken in the course of the Herschel-MESS programme. Both targets show spherically symmetric detached circumstellar emission, as is similarly seen around the stars presented in \citet{Kerschbaum2010}. The circumstellar environment of \object{S Sct} is rather well studied, especially in CO line emission. \citet{Olofsson1990} observed \object{S~Sct} with SEST and found by modelling a shell extended between \mbox{$\sim4-5.5\times10^{17}$\,cm} (adopting their assumed distance of 520\,pc, corresponding to $\sim50-70\arcsec$) from the pronounced double-peaked line profile. In subsequent studies \citep[e.g.][]{Olofsson1992,Bergman1993,Eriksson1993,Yamamura1993,Groenewegen1994} the shell mass and dynamical time-scales were determined. \object{RT~Cap,} on the other hand, was not known so far to have a similar detached structure. After \citet{Olofsson1988} found unresolved CO line emission, \citet{Schoier2001} modelled the CO observations with a present-day mass-loss at a rate of $1.0\times10^{-7}\,\mathrm{M_{\sun}\,yr}^{-1}$ and a corresponding envelope radius of $2.3\times10^{16}$\,cm (3\farcs 4), when adopting a distance of 450\,pc. With the high spatial resolution of the \textit{Photodetector Array Camera
and Spectrograph} (PACS) instrument \citep{Poglitsch2010} a highly detailed morphological analysis of the cold detached dust structures is now possible. 

\section{Observations}
\subsection{Herschel observations}
\object{S Sct} and \object{RT~Cap} (see Table~\ref{tab:targets} for basic properties) were observed in the course of the \textit{Herschel} Mass-loss of Evolved StarS (MESS\footnote{http://www.univie.ac.at/space/MESS/}) guaranteed time key programme \citep{Groenewegen2011} and are part of a larger, morphologically very heterogeneous sample of 78 targets \citep[see][]{Cox2012}. The observations were carried out using the photometer unit of PACS. The instrument provides imaging in three bands, of which we used two,  the 70\,$\mu$m (blue) and 160\,$\mu$m (red) band. The pixel scales of the detectors are 3\farcs 2 and 6\farcs 4, providing sufficient sampling for the corresponding nominal point-spread function (PSF) widths of 5\farcs 6 and 11\farcs 4. 
\object{S Sct} was observed twice, on 19-04-2011 (OBSID 1342219068 and 1342219069) and 20-09-2011 (OBSID 1342229077 and 1342229078), with a duration of 2574 and 2390 seconds.
\object{RT~Cap} was observed on 09-05-2010 (OBSID 1342196036 and 1342196037) for 2180 seconds, and was re-observed on 18-10-2011 (OBSID 1342231101 and 1342231102) for 15043 seconds. All observations were carried out in scanmap mode. In this setup the satellite is scanning the sky along parallel lines, usually repeating the pattern to improve the coverage. Each observation consists of two scans that are oriented perpendicular to each other.
The observational data were reduced and processed using the \emph{Herschel Interactive Processing Environment} (HIPE) with a modified pipeline script. For the mapping the \emph{Scanamorphos} code \citep{Roussel2013} was chosen because in our experience it provides the best results for the presented type of objects. It effectively removes detector drifts, making use of the highly redundant scanmap data, and hence best recovers the faint extended emission. In the final projection the two joint scans were mapped with a resolution of $1\arcsec$ and $2\arcsec$ per pixel in the blue and red bands. By this we significantly oversampled the PSF width mentioned above. Details about the data processing can be found in \citet{Groenewegen2011} and \citet{Ottensamer2011}. From the maps we extracted two-band photometry, with the numbers summarised in Table~\ref{tab:pacsphoto}.


\begin{table}
\centering
\caption{Target parameters.}
\label{tab:targets}
\begin{tabular}{lcccccc}
\hline
\hline
\vspace{-9pt}\\
Target & Spec. type & $P$ & $D$ & $T_{\mathrm{eff}}$ & C/O & $A_V$ \\
 & & [days] & [pc] & [K] & & [mag]\\
\hline
\vspace{-8pt}\\
S Sct & C6.4(N3) & 148 & 386 & 2755 & 1.07 & 0.3\\
RT~Cap & C6.4(N3) & 393 & 291 & 2480 & 1.10 & 0.2\\
\hline
\end{tabular}
\tablefoot{Variability and spectral data are taken from the GCVS \citep{Samus2012}, distances from Hipparcos parallaxes \citep{Leeuwen2007}. Effective temperature and C/O are taken from \citet{Lambert1986}. See text for the determination of $A_V$.}
\end{table}
\begin{table}
\centering
\caption{PACS aperture photometry measurements. The typical total error is 5\% and 10\% in the blue and red band. The rightmost column gives the radii of the applied apertures.}
\label{tab:pacsphoto}
\begin{tabular}{lccccc}
\hline
\hline
 & \multicolumn{2}{c}{$70\,\mu$m} & \multicolumn{2}{c}{$160\,\mu$m} &\\
 & total & shell & total & shell & shell$_\mathrm{in}$-shell$_\mathrm{out}$\\
 & [Jy] & [Jy] & [Jy] & [Jy] & [$\arcsec$]\\
\hline
S Sct & 11.6 & 6.1 & 5.1 & 3.1 & 55-80\\
RT~Cap & 3.9 & 0.8 & 1.6 & 0.6 & 80-110\\
\hline
\end{tabular} 
\end{table}

\begin{table}
\centering
\caption{References to photometric data used for the SED fitting.}
\label{tab:sed}
\begin{tabular}{lccc}
\hline
\hline
 & Opt. & NIR & MIR\,\&\,FIR\\
\hline
S~Sct & 1 & 2,3,4,5,6,7 & 9,10,11,12\\
RT~Cap & 1 & 5,6,7,8 & 9,10,11,12\\
\hline
\end{tabular}
\tablebib{(1)~\citet{Kharchenko2009}; (2)~\citet{Olofsson1993}; (3)~\citet{Noguchi1981}; (4)~\citet{Walker1980}; (5)~\citet{Kerschbaum1994}; (6)~\citet{Whitelock2006}; (7)~2MASS \citep{Cutri2003}; (8)~WISE \citep{Cutri2012}; (9)~AKARI/FIS \citep{Yamamura2010}; (10)~AKARI/IRC \citep{Ishihara2010}; (11)~IRAS PSC \citep{Beichman1988}; (12)~Herschel/PACS (this work, see Appendix A)}
\end{table}

\section{Modelling}
The modelling of the circumstellar dust distribution was made using a modified and extended version of the well-known radiative transfer code DUSTY \citep{Ivezic1999}, called \emph{MoD} (More of DUSTY), presented in \citet{Groenewegen2012}. In a nutshell, it employs DUSTY in a minimisation routine to determine the model that best fits the provided input data. Smooth, continuous envelopes and detached shells can be treated equally well.

\subsection{Model of a detached shell}
\begin{figure}[ht]
\centering
\resizebox{0.9\hsize}{!}{\includegraphics{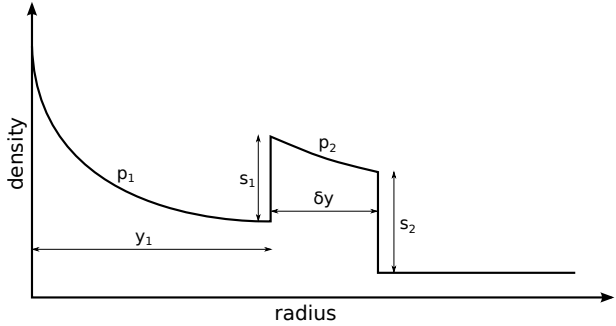}}
\caption{Schematic view of the parameters of the model density distribution. Dimensionless density is plotted against arbitrary radius. An explanation of the parameters is given in the text.}
\label{fig:modelscheme}
\end{figure}

The modelled envelope consists of three parts, realised by an extended form of a piece-wise power-law density distribution. The inner part represents the low mass-loss period after the detached-shell formation; its shape is described by the power-law parameter $p_1$. The second part is supposed to model the actual detached narrow shell that we can see, characterised by the parameters for size, $y_1$ and $\delta y$, and shape, $p_2$, $s_1$ and $s_2$. The latter two are scaling factors that introduce a sudden increase and drop in the density profile at the borders, reflecting the abrupt changes we assume in the mass-loss rate. For the outermost part the density is set to a low value (typically 0.1\% of the detached shell) and the size is chosen to extend somewhat beyond the detached part. A schematic representation is given in Fig.~\ref{fig:modelscheme}. As input, all of the above parameters are set to best-guess values, giving an initial density profile with which the modelling routine is fed. 

\emph{MoD} further requires information about the radiation source and the dust properties. For the radiation source (i.e. the central star) COMARCS model atmospheres \citep{Aringer2009} were adopted. The model with the appropriate parameters was selected based on effective temperature and C/O ratio values found in \citet{Lambert1986}, see Table~\ref{tab:targets}. For the other COMARCS parameters mass, metallicity and log g, reasonable values were assumed ($M$=2\,M$_\odot$, $Z$=Z$_\odot$, $log\,g$=-0.4). Since only the relative shape of the COMARCS spectrum is of interest for $MoD$, deviations of the latter three parameters can be neglected. 
As obtained from the online database, the models extend from a wavelength of 0.44\,$\mu$m to 25\,$\mu$m. To also cover short wavelengths, the Aringer models were extrapolated to 0.1\,$\mu$m assuming a black body at the respective temperature. For the long wavelength tail the DUSTY routine extrapolates the input spectrum according to Rayleigh-Jeans.
Since the studied objects are carbon stars, a mixture of amorphous carbon \citep[amC,][AC1 species]{Rouleau1991} and a small fraction of silicon carbide \citep[SiC,][]{Pitman2008} was selected as the grain constituents. The SiC was added because of the 11\,$\mu$m signature in the SWS spectrum of S~Sct and the RT~Cap LRS spectrum. 
The optical properties of the dust were set by adopting a distribution of hollow spheres \citep[DHS,][]{Min2003} with a highest vacuum-inclusion volume fraction of 0.7 and a grain radius of 0.15\,$\mu$m. This kind of grain geometry was found to describe the absorption and scattering properties of dust particles quite well \citep{Min2003,Min2005}. 

Furthermore, we provide an estimate for stellar luminosity $L$ and optical thickness $\tau$ of the whole dust envelope as initial guesses for the model. These two values and the density distribution parameters $y_1$, $\delta y$, $p_1$, $p_2$ and $s_1$ can each be set as variable or kept fixed for the fitting. 

\subsection{Observational constraints}
To fit the model spectrum, photometry from the new PACS observations and archive data (VRIJHKL, IRAS) were used (see Table~\ref{tab:sed} for references and Table~\ref{tab:photdata} for the used data+errors). For multiple observations per filter, all datapoints were included individually, regardless of the observation epoch. Hence variations due to the pulsational cycle average to a mean value. The photometric data were de-reddened using the interstellar extinction $A_V$, estimated from the stellar galactic position. For this, the procedure used in \citet{Groenewegen2008} was followed, which is based on the methods of \citet{Parenago1940}, \citet{Arenou1992}, \citet{Drimmel2003} and \citet{Marshall2006}. The adopted values are listed in Table~\ref{tab:targets}. Where available, ISO SWS spectra from \citet{Sloan2003} and IRAS LRS \citep{Volk1989} were used as an additional constraint for the spectral energy distribution (SED) fit. Additionally, the new Herschel/PACS observations provide spatial information, which is incorporated by azimuthally averaged radial profiles at 70 and 160\,$\mu$m. This constrains the radial brightness distribution of the model. Details on the Herschel PACS data can be found in Appendix~\ref{apdx1}.

\subsection{Modelling and output}
First, the luminosity was determined by fitting the model spectrum to the photometric measurements. The radial brightness profiles were omitted because the detached shell was neglected. Once $L$ was determined, it was kept fixed for the following fitting of the shell's shape parameters and $\tau$. For this, now the radial brightness profiles were taken into account as well. Out of the remaining six variables, four at most were allowed to be varied (without preset upper/lower bounds) at once in the minimisation during a run. As default, the power-law indices $p_1$ and $p_2$ of the density profile were kept at -2, which is what is theoretically expected in the ideal case of a perfectly spherically symmetric environment and a smooth wind expanding at constant velocity. While $y_1$, $\delta y$, $s_1$ and $\tau$ were variable, next the inner radius of the detached shell was determined and fixed later because it can be reliably fitted. If necessary, $p_1$ and $p_2$ were also allowed to vary to improve particularly the brightness profile of the model.


Each calculated fit was quantitatively evaluated by
\begin{equation}
\chi^2=\sum \limits_{i=1}^{n}(m_\mathrm{obs}(i)-m_\mathrm{pred}(i))^2/\sigma^2_\mathrm{m_{obs}(i)},
\end{equation}

which is provided by the model for the various datasets (photometry (see Table~\ref{tab:photdata}), spectroscopy and intensity profiles) separately and summed up as well. Furthermore, the goodness of fit was also assessed by penalising larger numbers of parameters that were kept free during minimisation ($p$), defined as
\begin{equation}
\mathrm{BIC}=\chi^2+(p+1)\,\ln(n),
\end{equation}
after the Bayesian information criterion \citep[see][]{Schwarz1978}, where $n$ is the number of measurements. For more details we refer to section~2 in \citet{Groenewegen2012}.

From the fitted model parameters the mass of the dust contained in the detached shell can be calculated using the relation
\begin{equation}
M_\mathrm{dust}=\frac{16\,\pi^2\,a\,\rho\,\tau_{0.55}\,\tilde{r}^{\,2}}{3\,Q_\mathrm{abs,0.55}},
\label{eq:mass}
\end{equation}
where \emph{a} is the grain radius, $\rho$ the bulk density and $Q_\mathrm{abs}$ the absorption efficiency of the chosen grain properties. $\tau_{0.55}$ is the optical depth of only the detached shell at 550\,nm as obtained from the best model fit, and $\tilde{r}$ its radius (i.e. the radial centre of mass). For spherically symmetric density distributions following a power law $\rho\propto r^{-p}$, the shell's $\bar{r}$ can be determined using the derived formula
\begin{equation}
\tilde{r}=\begin{cases}
   2^{^1/_{p-3}} \left(y_1^{3-p}+y_2^{3-p}\right)^{^1/_{3-p}} & \text{if } p \ne 3 \\
   \sqrt{y_1 y_2}                                                 & \text{if } p = 3
  \end{cases}.
\end{equation}
$y_1$ and $y_2$ are the inner and outer boundaries of the detached shell (where $y_2=y1+\delta y$), determined by the best-fit model.

In view of the large parameter space and possible observed deviations from model assumptions (see Sect.~\ref{results}), it is not always straightforward to find an objectively best fit in the sense of the globally lowest $\chi^2$ or \emph{BIC} value possible. It can even happen that models with a low overall error obviously do not match the observations, especially the brightness profile. Therefore, instead of relying on a single fit, several models were calculated with different weights on spectral, photometric and spatial brightness information and changed dust properties. However, this further increase of degrees of freedom leads to an even more diverse sample of fitted parameter sets, which makes it virtually impossible to objectively pin down a single model as the best result. Therefore, the most suitable set of best-fit models (corresponding to a particular error weighting) was chosen subjectively by eye and from that subset the model with the lowest $\chi^2$ (combined error from taking into account photometric, spectroscopic and brightness profile information) was selected.

%
%
%
%
%

\section{Results}
\label{results}
\begin{table}
\centering
\caption{Kinematic information for the target stars in the heliocentric reference system (H) and corrected for the solar motion (G). $\theta$ is the inclination of the motion vector with respect to the plane of sky. Proper motion and parallax data are taken
from \citet{Leeuwen2007}.}
\label{tab:kinematics}
\begin{tabular}{lccccc}
\hline
\hline
&$\mu$ [mas\,yr$^{-1}$]&P.A.&$v_\mathrm{r}$ [km\,s$^{-1}$]&$v_\mathrm{space}$ [km\,s$^{-1}$]&$\theta$\\
\hline
\vspace{-8pt}\\
\multicolumn{2}{l}{S~Sct} & & & & \\
H& 9.1 & $119.5^\circ$ & -0.2 & $16.7\pm3.8$ & $-0.7^\circ$\\
G& 6.6 & $78.0^\circ$ & 12.9 & $17.7\pm2.3$ & $47^\circ$\\
\hline
\vspace{-8pt}\\
\multicolumn{2}{l}{RT~Cap} & & & & \\
H& 13.3 & $156.0^\circ$ & $-29.2$ & $34.4\pm2.6$ & $-58^\circ$\\
G& 2.9 & $185.7^\circ$ & $-20.9$ & $21.2\pm2.4$ & $-79^\circ$ \\
\hline
\end{tabular}
\end{table}
\subsection{\object{S~Sct}}
\begin{figure}$
\begin{array}{c}
\includegraphics[width=0.9\hsize]{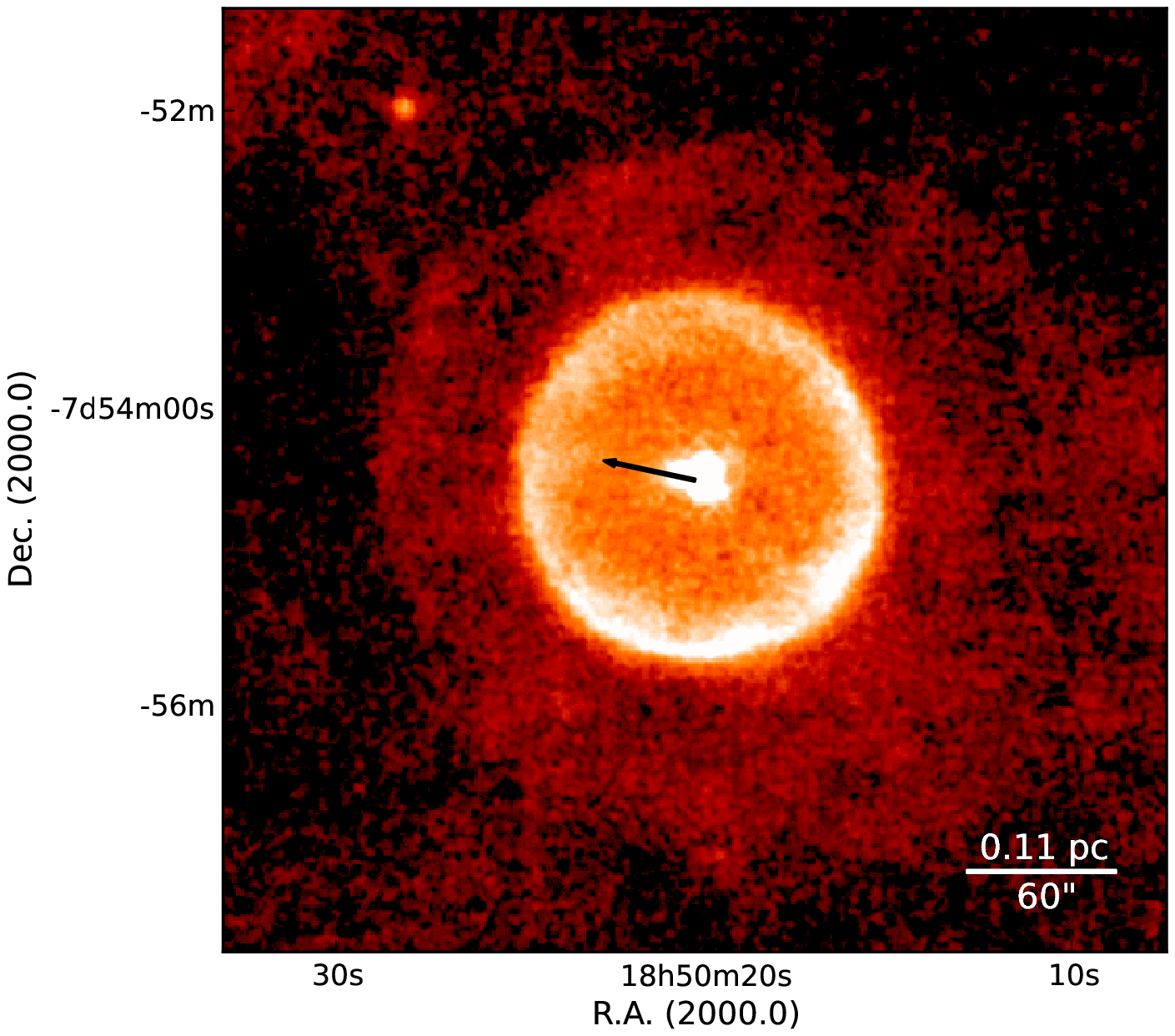} \\
\includegraphics[width=0.9\hsize]{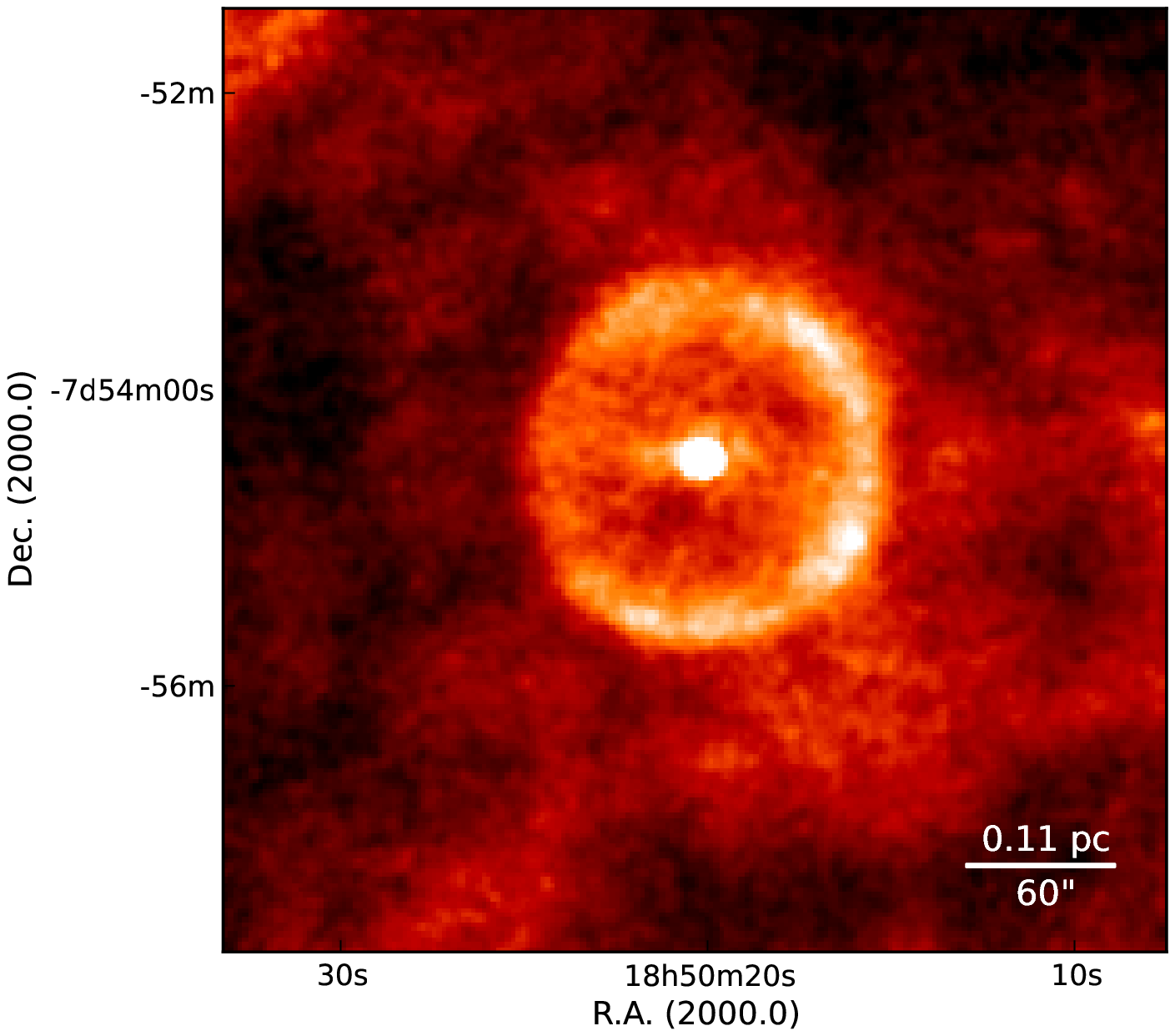}
\end{array}$
\caption{\object{S Sct} as seen by PACS in in the 70\,$\mu$m (top) and 160\,$\mu$m (bottom) band. The arrow in the top image indicates the proper motion (corrected for solar motion), its length shows the distance travelled in $\sim$5800 years. The spatial resolution (FWHM) is 5\farcs 8 and 11\farcs 5, measured from point sources present in the field.}
\label{fig:ssct70}
\end{figure}

\subsubsection{Large-scale morphology -- Herschel data}
\object{S Sct} exhibits a very bright spherically symmetric detached circumstellar shell in both PACS bands (Fig.~\ref{fig:ssct70}). The brightness peaks at $69\arcsec$ from the star, which corresponds to a linear value of 0.13\,pc, adopting a distance of 386\,pc. Additionally, a so far unknown structure is detected at a distance of approximately $130\arcsec$ (0.25\,pc). The proper motion is headed towards a P.A. of $78^\circ$ (see Table~\ref{tab:kinematics}) after transformation to the galactic reference system \citep[as outlined by][]{Johnson1987}. Together with the inclination angle of 47$^\circ$ between the plane of sky and the space motion vector,
this suggests that the outer shape is influenced by wind-ISM interaction. The effect, however, is expected to be weak because
of the low space velocity of the star (17.7\,km\,s$^{-1}$) with respect to the LSR, which is assumed to be co-moving with the ISM material around \object{S Sct}. If we assume that the ISM has not slowed down the stellar wind significantly (adopting a generic value of 15\,km\,s$^{-1}$), the estimated dynamical age is about 16\,000 years. 

\begin{figure*}
\includegraphics[width=\hsize,clip]{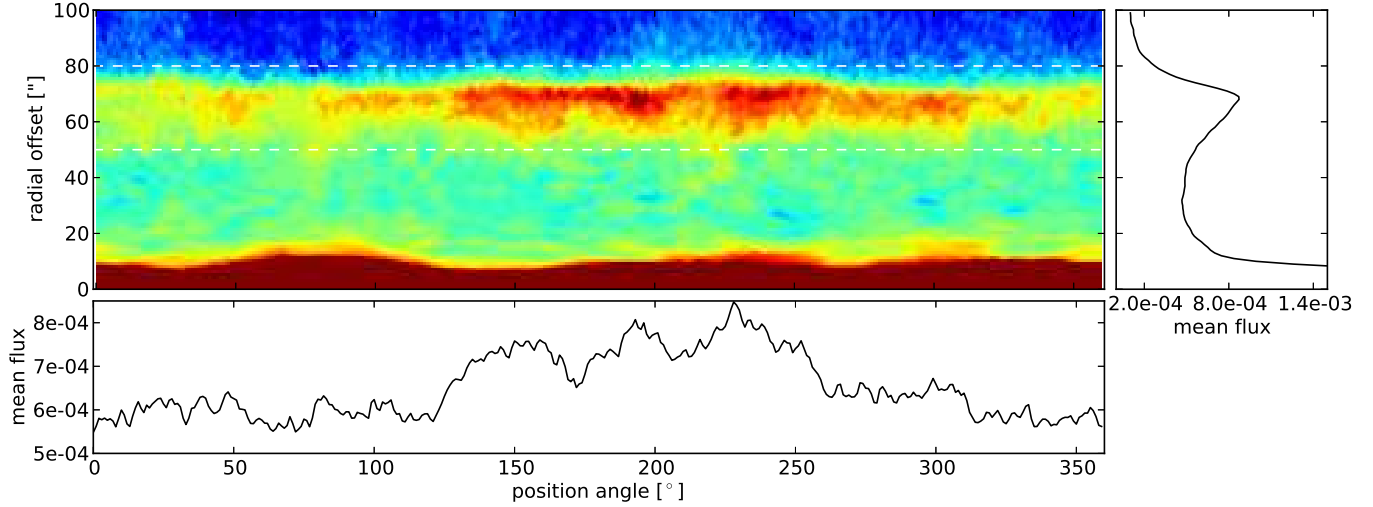}
\caption{\object{S~Sct} 70\,$\mu$m data mapped onto polar coordinates. The right graph shows the azimuthally (horizontally) averaged flux, the bottom profile the radial (vertical) average of the shell brightness between 50\arcsec and 80\arcsec offset (indicated by the dashed lines). The mean flux is given in units of \mbox{Jy arcsec$^{-2}$}.}
\label{fig:ssct_polar}
\end{figure*}

The centre of the inner shell coincides with the stellar position, meaning that we cannot see a displacement within the accuracy of the $70\,\mu$m map ($\approx\!1\arcsec$), which could potentially occur because of the space motion. While there are no major deviations from the overall ``roundness'' of the shell, the brightness is not evenly distributed azimuthally. This becomes even more evident when the observation is mapped onto polar coordinates (Fig.~\ref{fig:ssct_polar}). In the north-east part the shell is weak by comparison, most of the emission is observed at the opposite side. 
The 160\,$\mu$m data show a similar brightness distribution, only the north-west part is more prominent here, compared to the shorter wavelength.

\subsubsection{Model of the detached shell}
\begin{figure}
\includegraphics[width=0.9\hsize]{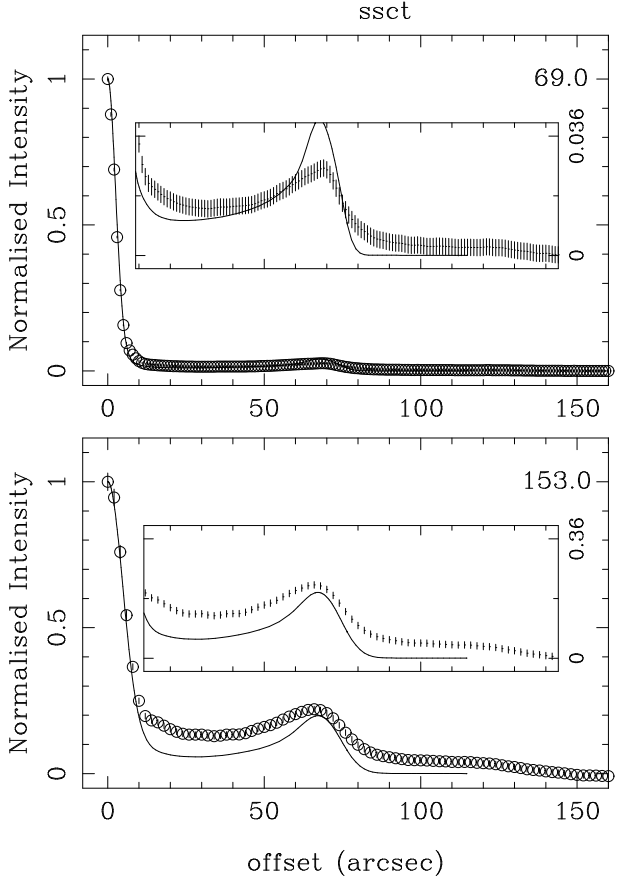}
\caption{Radial intensity profiles for the model fit (solid line) compared with PACS observations at 70\,$\mu$m (top) and 160\,$\mu$m (bottom). The inset (x-scale aligned) shows a vertical magnification of the detached-shell region.}
\label{fig:ssct_int}
\end{figure}
\begin{figure}
\includegraphics[width=0.9\hsize]{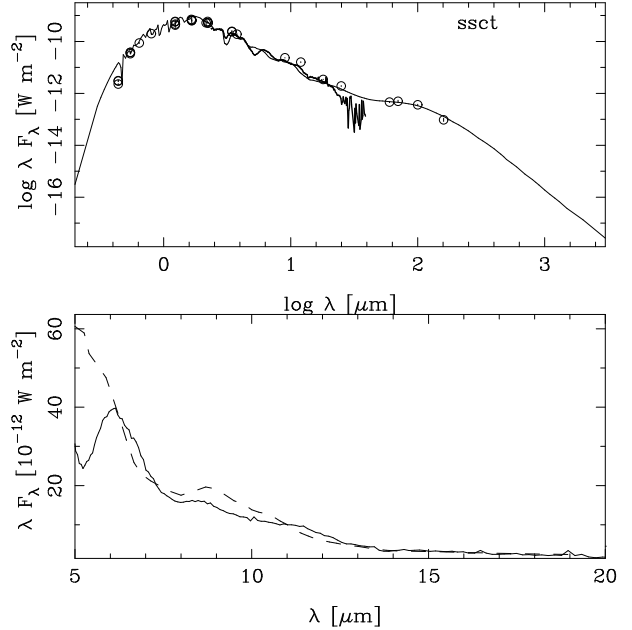}
\caption{Top: SED of the model fit and photometric data for \object{S Sct}, including an SWS spectrum (for references see Table\,\ref{tab:sed}). Bottom: magnified region around the SiC feature, comparing the model (dashed) with the SWS spectrum.}
\label{fig:ssct_sed}
\end{figure}
The adopted COMARCS model has a $T_\mathrm{eff}$ of 2900\,K and a C/O ratio of 1.05 (see Table~\ref{tab:targets}). For the dust a mixture of 95\% amC and 5\% SiC (due to the present 11\,$\mu$m feature) was selected, with a bulk density of 1.34\,g\,cm$^{-3}$. 

While the SED, including the spectra, were approximated quite well in all wavelength regions (see Fig.~\ref{fig:ssct_sed}), it was more difficult to obtain a model that satisfyingly reproduces the radial intensity distribution, even when adjusting the weight given to its datapoints (Fig.~\ref{fig:ssct_int}). In the end, models obtained with scaled-down errorbars for the radial brightness distribution (by a factor of up to 4) yielded slightly better overall results because the SED fit does not deteriorate significantly. They were thus used to derive the shell parameters.
Typically, the calculations underestimate the brightness and hence the dust mass that is due to the more recent mass-loss. This flaw is seen in models for several objects and is not restricted to \object{S Sct} alone \citep[cf. \object{TT~Cyg} in][]{Groenewegen2012}. A possible explanation is a gradual decrease in the mass-loss rate after the high mass-loss phase (as was found by \citet{Maercker2012} for R~Scl) instead of the abrupt change that is assumed in the model.
In the ideal case of a perfectly spherically symmetric environment and a smooth wind expanding at constant velocity, the density is assumed to be distributed as $r^{-2}$. The observation, however, suggests deviating gradients (flatter) in the density drop. Our obtained $p_1$ value of $1.0\pm0.5$ indicates that the rarefication due to expansion is partly compensated for by a gradually decreasing mass-loss rate (MLR) after (i.e. inside) the detached shell. In general, but particularly in this case, $p_2$ cannot be constrained well and lies between typically 1 and 6, in this case it was therefore fixed to 2. The fit to the radial profile was slightly improved by giving the brightness profile more weight, that is, scaling down the errors, while the weight of the ISO spectrum was reduced. A summary of the model fit parameters is given in Table~\ref{tab:model}. 
Some of the fit parameters are found to be more correlated to some degree. A strong dependence, or linear (Pearson) correlation ``cor'', exists between the shell thickness  and the density contrast (cor$(\delta y,s_1)\approx-0.9$). A correlation of 0.8 is found for $y$ and $s_1$, 0.7 between $\tau$ and $s_1$ and -0.7 for $y$ and $\delta y$.
Despite the rather poor fit to the brightness distribution, including that data set still improves the information on the density distribution (e.g. the increase in mass-loss rate) and adds certainty to the detached-shell position and thus the dust temperature.

For \object{S Sct} $T_\mathrm{d}$ at the inner boundary of the detached shell ($66\arcsec\hat{=}0.13$\,pc) is $46$\,K and drops to $42$\,K at the outer edge ($84\arcsec\hat{=}0.17$\,pc). The calculated formal error is small ($\sim1$\,K) because some factors
that were difficult to estimate were not taken into account (see Sect.~\ref{errordisc}). Based on the model parameters, we also calculate ${(7\pm2)}\times10^{-5}\,\mathrm{M_{\sun}}$ for the dust mass contained in the detached shell using Eq.~\ref{eq:mass}. Adopting the gas expansion velocity of 16.5 km\,s$^{-1}$ found by \citet{Schoier2005} also for the dust component (that is, assuming no drift), the epoch of enhanced mass-loss would have lasted about at least 1000 years. This would correspond to an average dust mass-loss rate of $6\times10^{-8}\,\mathrm{M_{\sun}}$\,yr$^{-1}$ during the formation of the detached shell. Assuming a gas-to-dust mass ratio of 200, the total MLR would be $1.2\times10^{-5}\,\mathrm{M_{\sun}}$\,yr$^{-1}$ during that phase.


\subsection{\object{RT~Cap}}
\subsubsection{Large-scale morphology -- Herschel data}
\label{rtcap_morph}
\begin{figure*}$
\begin{array}{cc}
\includegraphics[width=0.45\hsize]{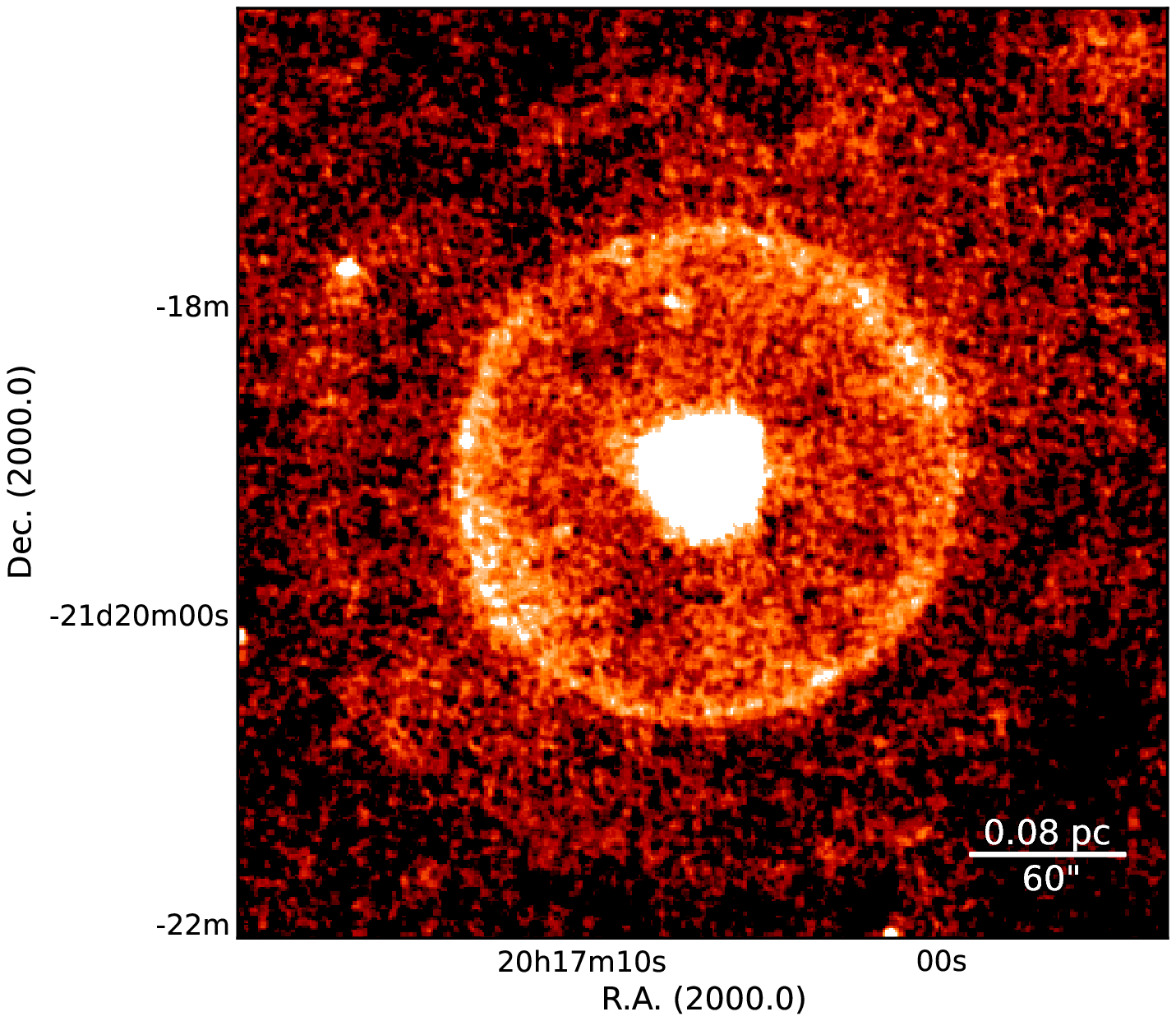} &
\includegraphics[width=0.45\hsize]{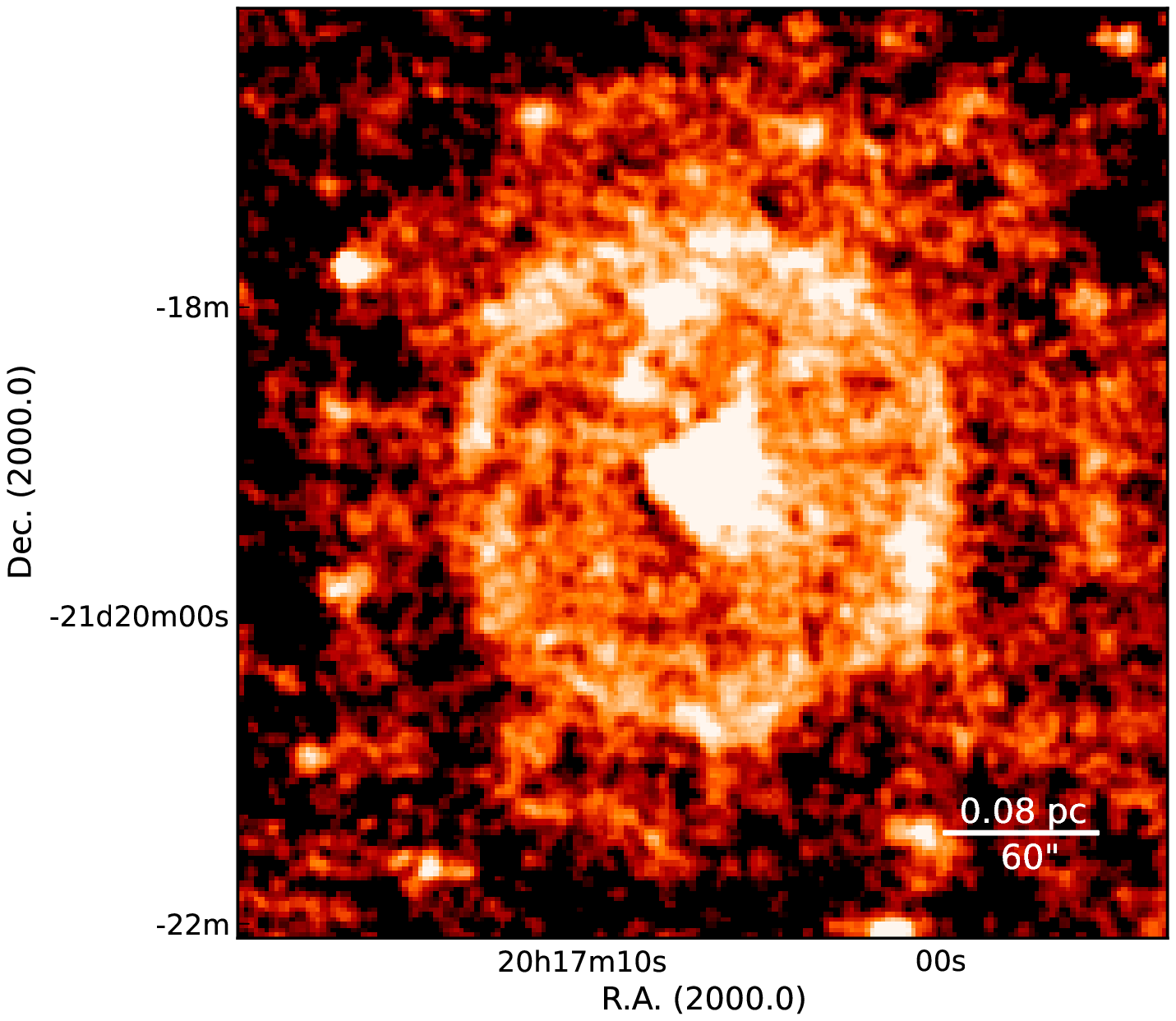}
\end{array}$
\caption{PACS maps of \object{RT~Cap} at 70\,$\mu$m (left) and 160\,$\mu$m (right). The spatial resolution (FWHM) is 5\farcs 8 and 11\farcs 5.}
\label{fig:rtcap70}
\end{figure*}

\begin{figure*}
\includegraphics[width=\hsize]{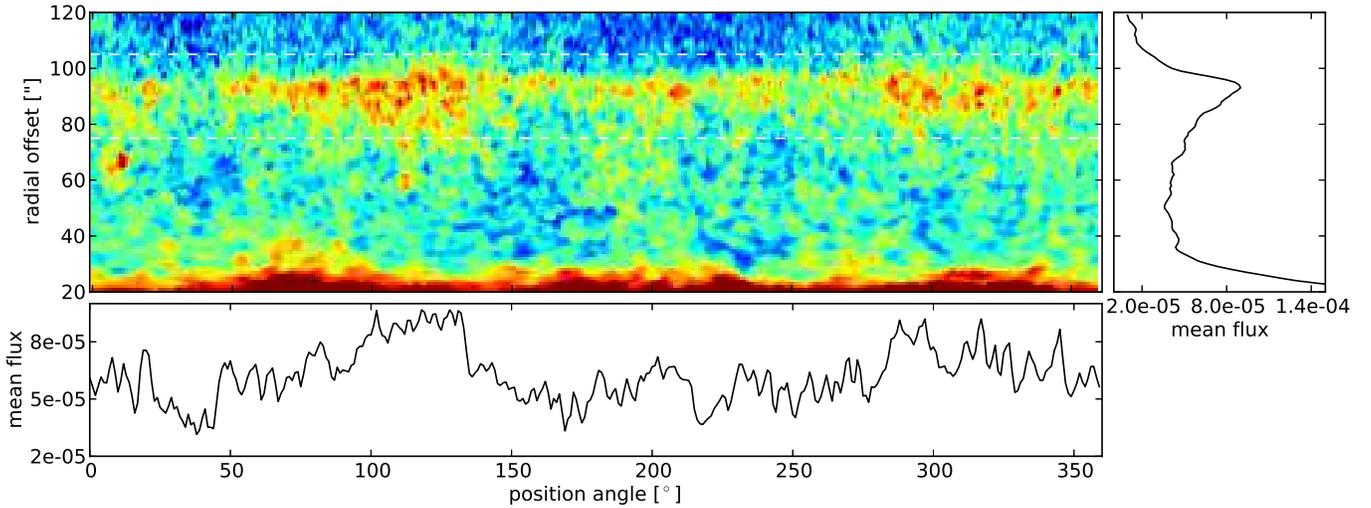}
\caption{Representation of \object{RT~Cap} 70\,$\mu$m data in polar coordinates. The radial average of the shell (bottom profile) was taken between 75\arcsec and 105\arcsec (dashed lines). Mean flux is given in Jy arcsec$^{-2}$.}
\label{fig:rtcap_polar}
\end{figure*}

Clearly visible in both PACS bands, \object{RT~Cap} is surrounded by a remarkably thin, spherically symmetric detached shell of dust (Fig.~\ref{fig:rtcap70}). The envelope brightness peaks at a radial offset of 94$\arcsec$ from the star, corresponding to 0.13\,pc at the adopted distance of 291\,pc. Again, no offset in RA or DEC between star and shell centre is measurable within the accuracy of the PACS data. 
As Fig.~\ref{fig:rtcap_polar} clearly shows, the dust is not distributed smoothly, but exhibits a very patchy structure. Moreover, the observed brightness is increased in the eastern and western shell sections in the 70\,$\mu$m data. In the 160\,$\mu$m map, the bright regions are located differently, maybe indicating an azimuthally varying temperature distribution. While there are several structures that we can clearly resolve, parts of the detached shell seem thinner than our resolution limit. In
addition to the shell at 94$\arcsec$ , we detect another even fainter structure farther out. It is also roughly circularly shaped with a radius of $\sim155\arcsec$ (0.22\,pc) with only a weak signature in the azimuthally averaged radial profile.



\subsubsection{Model of the detached shell}
\begin{figure}
\includegraphics[width=0.9\hsize]{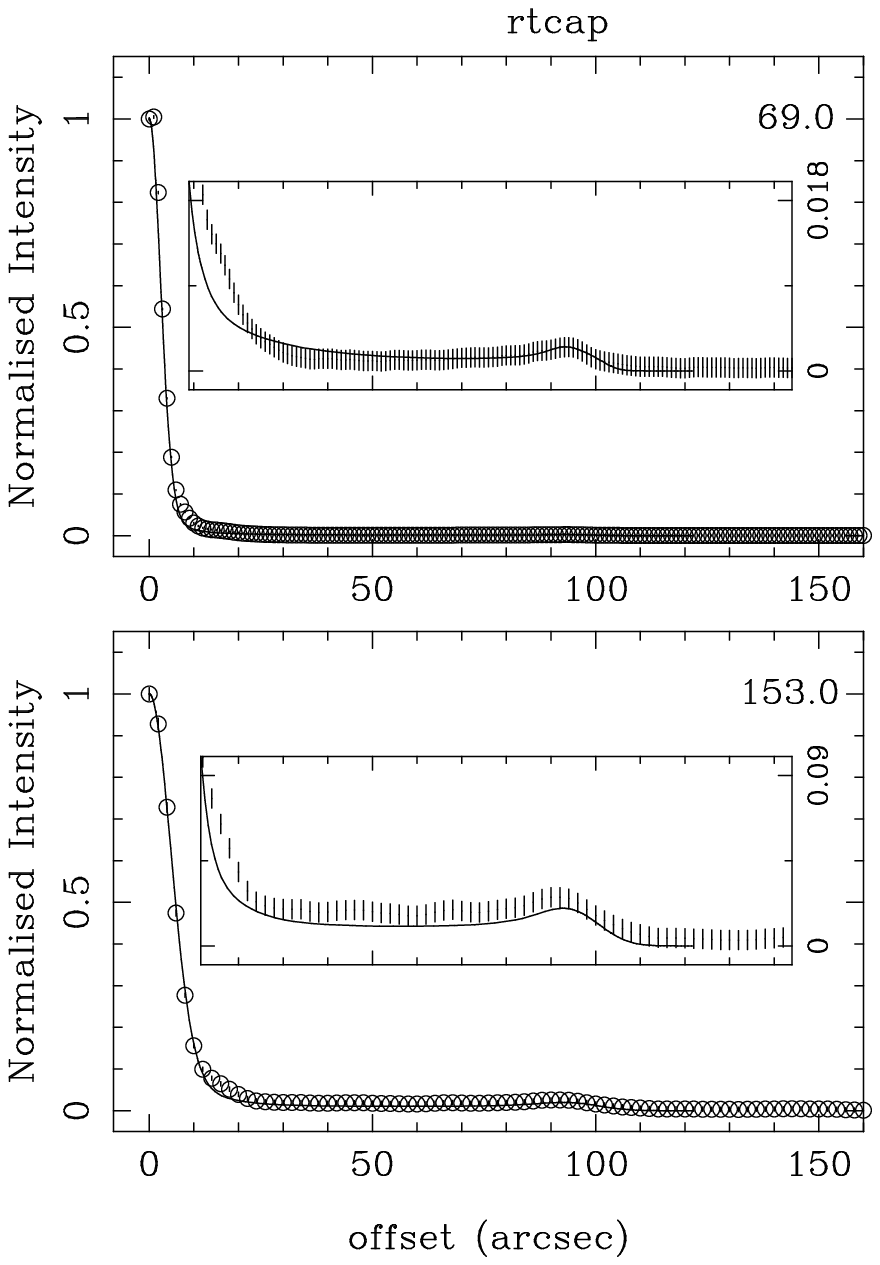}
\caption{Model fit to azimuthally averaged radial profiles of the PACS observations of \object{RT~Cap} at 70\,$\mu$m (top) and 160\,$\mu$m (bottom).}
\label{fig:rtcap_int}
\end{figure}

\begin{figure}
\includegraphics[width=0.9\hsize]{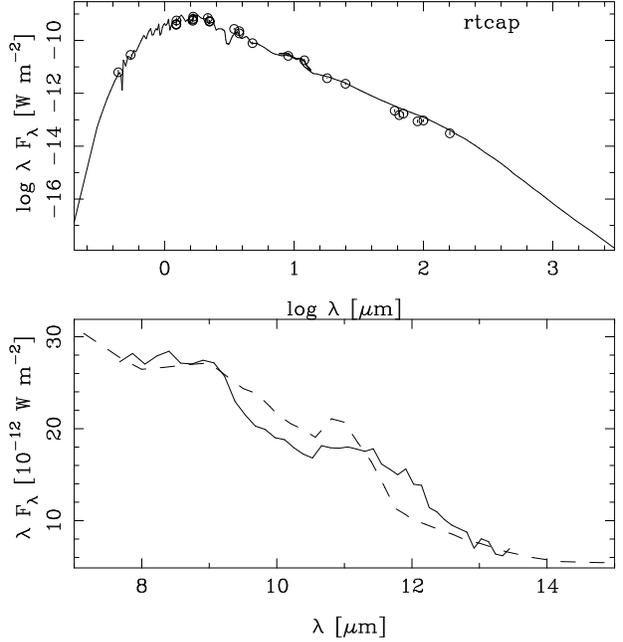}
\caption{Top: SED of the \object{RT~Cap} model fit compared with the photometric data (for references see Table\,\ref{tab:sed}), part of an LRS spectrum was also taken into account. Bottom: comparison between the LRS (solid line) and the model spectrum. The SiC feature in the model is narrower than observed.}
\label{fig:rtcap_sed}
\end{figure}

For the modelling, the outer, very faint structure was not considered at first. 
For the COMARCS spectrum a C/O ratio of 1.10 and $T_\mathrm{eff}=2500$\,K was adopted. 
The grain properties are the same as for \object{S~Sct}. 
In the resulting model (Figs.~\ref{fig:rtcap_int} \& \ref{fig:rtcap_sed}) the inner shell radius is located at 93$\arcsec$ (0.13\,pc), where the dust temperature is $39$\,K. The shell thickness is 7$\arcsec$ and the temperature at the outer boundary is $38$\,K. We compute the dust mass in the detached shell to be $(1.0\pm0.2)\times10^{-5}\,\mathrm{M_{\sun}}$.

For several pairs of varied parameters some degree of correlation can be found. The strongest degeneracy exists between the shell thickness  and the density contrast (cor$(\delta y,s_1)\approx-0.9$). A correlation of 0.8 is found for $y$ and $s_1$, 0.7 between $\tau$ and $s_1$ and $-0.7$ for $y$ and $\delta y$.

\begin{figure}
\resizebox{\hsize}{!}{\includegraphics[trim=0 0 0 10,clip]{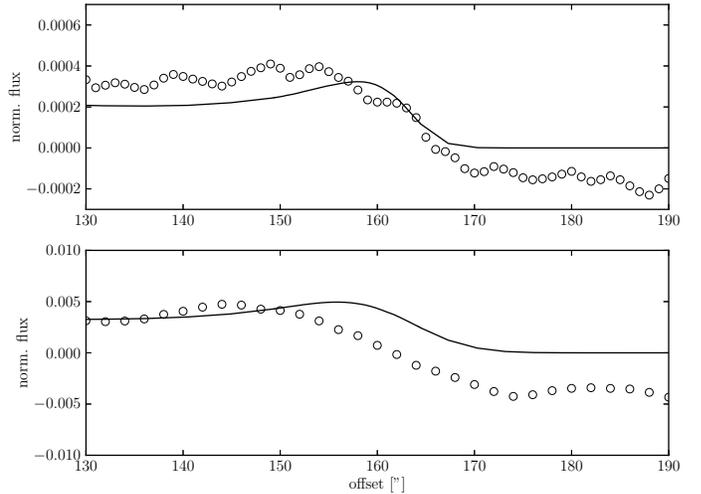}}
\caption{Model brightness fit (solid line) to the outer shell
of \object{RT~Cap}  at 70\,$\mu$m (top) and 160\,$\mu$m (bottom).}
\label{fig:rtcap_outer}
\end{figure}

A model of the outer structure at $\sim0.22$\,pc from the star (see Fig.~\ref{fig:rtcap_outer}) yields 32\,K compared with $37\pm3$\,K estimated from the flux ratio. From the poor fit we also obtain a dust mass of roughly $6\times10^{-6}\,\mathrm{M}_{\sun}$ for the structure.

There are no detections of a detached molecular shell around \object{RT~Cap}, therefore we adopted a generic value of 15\,km\,s$^{-1}$ for the expansion velocity of the ejected matter. This is probably a more appropriate value than adopting the 8\,km\,s$^{-1}$ measured for the present-day mass-loss because velocities of current winds tend to be considerably lower than those from past high mass-loss events \citep[see, e.g., Table~6 in][]{Schoier2001}. The assumed speed yields a duration of enhanced mass-loss of approximately 700--900 years, corresponding to an average dust mass-loss rate of $(2.3\pm0.8)\times10^{-8}\,\mathrm{M_{\sun}}$\,yr$^{-1}$ for the total dust mass of $(1.0\pm0.2)\times10^{-5}\,\mathrm{M_{\sun}}$ contained in the detached shell. Adopting a gas-to-dust ratio of 200, the total MLR is $5\times10^{-8}\,\mathrm{M_{\sun}}$\,yr$^{-1}$. However, considering that the shell's width is not constant around the position angle and the azimuthal average is broadened by the extended patches, the derived time is very likely an overestimate. Therefore, the actual formation duration is probably lower, leading to a higher MLR than given. Unaffected by this problem, we estimate the dynamical age of the enhanced mass-loss event to be 8500 years, again assuming an expansion velocity of 15\,km\,s$^{-1}$. 



\begin{table*}
\centering
\caption{Summary of the obtained model parameters plus statistical modelling errors.}
\label{tab:model}
\begin{tabular}{lrrcrrrrrc}
\hline
\hline
\vspace{-8pt}\\
 & $L\,[\mathrm{L_{\sun}}]$ & $\tau_{0.55\mu\mathrm{m}}$ & $T_\mathrm{d}\,[\mathrm{K}]$ & $p_1$ & $y_1\,[\mathrm{pc}]$ & $\delta y\,[\mathrm{pc}]$ & $s_1$ & $p_2$ & $M_\mathrm{d}\,[\mathrm{M_{\sun}}]$\\
\hline
\vspace{-8pt}\\
S~Sct & $4300\pm100$ & $0.03\pm0.003$ & $46$ & $1.0\pm0.5$ & $0.13\pm0.01$ & $0.02$ & $45\pm10$ & $2.0$ & $(7\pm2)\times10^{-5}$\\
RT~Cap & $2200\pm50$ & $0.21\pm0.05$ & $39$ & $1.75\pm0.03$ & $0.13\pm0.01$ & $0.016\pm0.007$ & $13\pm6$ & $2.0$ & $(1.0\pm0.2)\times10^{-5}$\\
\hline
\end{tabular} 
\tablefoot{$L$ is the luminosity of the central star, $\tau$ is the optical depth of the entire dust envelope. $y_1$ and $\delta y$ are the radius and width of the detached shell, given in pc. The condensation temperature $T_\mathrm{c}$ was kept fixed at 1000\,K. $T_\mathrm{d}$ is the dust temperature at distance $y_1$ and $M_\mathrm{d}$ is the dust mass in the detached shell. The given absolute shell thicknesses $\delta y$ correspond to an angular size of 10$\arcsec$ and 11$\arcsec$ for S~Sct and RT~Cap. The provided errors are the variation of the parameters that would yield a $\chi^2=1$, including all observational constraints. For the shell thickness of S~Sct no error is provided, since due to large variations of $\delta y$ no credible value could be obtained and the parameter was eventually kept fixed.}
\end{table*}

\section{Discussion}
\subsection{\object{S~Sct}}
\label{disc:ssct}
In the early 1990s, CO line emission was investigated in a number of papers. \citet{Olofsson1990} observed \object{S~Sct} with SEST and found by modelling a shell extended between \mbox{$\sim4$-$5.5\times10^{17}$\,cm} (adopting their assumed distance of 520\,pc, corresponding to $\sim50$-$70\arcsec$) from the pronounced double-peaked line profile. CO maps from \citet{Olofsson1992} confirmed these values. 
From our PACS measurements, we determine the radius of the detached dust structure to $3.9\times10^{17}$\,cm, which suggests that the gas and dust component are roughly co-spatial.  

From detailed modelling \citet{Eriksson1993} found that the formation period of the shell was 10\,000 years ago and lasted about 400 years. Adopting a gas expansion velocity of 16.5 km\,s$^{-1}$ \citep{Schoier2005} for the dust component as well (that is, assuming no velocity drift), the epoch of enhanced mass-loss would have lasted about 1000 years in our case. Since the detached-shell feature in the azimuthally averaged PACS intensity profile is broadened to some extent because of the deviations from spherical symmetry (see Fig.~\ref{fig:ssct_polar}), the true shell thickness may be slightly overestimated by the model (which gives $10\arcsec\hat{=}0.02$\,pc), and likewise, the duration of the high-mass-loss period. However, given the quality of the model fit to the brightness distribution (Fig.~\ref{fig:ssct_int}), this factor is not expected to have a significant impact on the derived parameters. Furthermore, the spatial resolution of the 70\,$\mu$m PACS maps ($5\farcs6$ PSF) is not expected to limit the determination of the shell width (10\arcsec). That our obtained shell thickness (or formation time) nevertheless does not compare favourably with the findings from molecular line data or polarised scattered-light measurements \citep[100-200\,yrs, e.g.][]{Ramstedt2011} can be explained if we assume a dispersion in the expansion velocity of the dust grains. Adopting a dispersion of only 1\,km\,s$^{-1}$ would lead to an additional spread mimicking almost 500 years of enhanced mass-loss after 7500 years of expansion. Because the shells observed in scattered light are smaller, that is, younger, than those presented in this paper, the broadening due to velocity dispersion would be lower there. Considering this effect, the 1000 years we derive for the duration of high MLR are likely an overestimate and the actual time of enhanced mass-loss is indeed of the order of some 100 years. However, complementary data at different wavelengths would help to ensure that the dust is indeed as broadly distributed as it appears from our PACS data.

\citet{Schoier2005} investigated several objects known to have detached shells. Modelling both components of the circumstellar envelope of S~Sct, they determined shell masses of $(7.5\pm1.0)\times10^{-3}\,\mathrm{M_{\sun}}$ and $(8.0\pm7.0)\times10^{-4}\,\mathrm{M_{\sun}}$ for the gas and dust, respectively. They also derived a dust temperature of $34\pm9$\,K.
Similar to their estimate of a corresponding dynamical age of the shell of 8100 years, we obtain $\sim$7500 years for the adopted distance of 400 pc
they used. 
Comparing our obtained dust mass with the gas mass estimates from previous molecular line observations, the gas-to-dust ratio of the detached shell lies in the plausible range between approximately 100 and 300, considering \citet{Schoier2005} and \citet{Eriksson1993}, respectively.

Modelling by \citet{Bergman1993} showed that the observed patchy CO distribution is best explained by the fact that the gas is not smoothly distributed throughout the shell, but is instead concentrated in about $10^3$ clumps, $1.9\times10^{16}$\,cm (2\farcs 4) in radius. While we also see minor small-scale brightness variations in our PACS maps, the spatial resolution is too low to actually identify such clumps. 
In a study of resolved CO ($J$=$1\!\!\rightarrow\!\!0$) data from the 45 m radio telescope at Nobeyama Radio Observatory \citet{Yamamura1993} found that during the formation of the detached shell \object{S~Sct} lost mass at a rate of $\ge 1.3\times10^{-4}\,\mathrm{M_{\sun}}$\,yr$^{-1}$, assuming a smooth density distribution. This is considerably higher than the few $10^{-5}\,\mathrm{M_{\sun}\,yr^{-1}}$ that are needed to explain the observations using a clumped medium, according to \citet{Bergman1993}.



\subsection{RT Cap}
\object{RT~Cap} has been observed in CO \citep{Olofsson1988}, but no indications of a detached envelope were found. In a study of molecular radio-line emission \citet{Schoier2001} modelled the CO signal with a present-day mass-loss at a rate of $1.0\times10^{-7}\,\mathrm{M_{\sun}\,yr}^{-1}$, where the matter is expelled at 8\,km\,s$^{-1}$. For the corresponding envelope size they derived a radius of $2.3\times10^{16}$\,cm (3\farcs 4), when adopting a distance of 450\,pc. Resolving a dusty counterpart is clearly out of range for the Herschel-PACS observations.


The physical diameter of the newly found dust shell around \object{RT~Cap} is very similar to that surrounding \object{S~Sct} ($\sim4\times10^{17}$\,cm $\approx0.13$\,pc). The probability of also detecting a corresponding CO shell around the former is very low, however. Because of the very low MLR, most of the ejected CO has probably been dissociated at the distant radii. The space motion of 21.2\,km\,s$^{-1}$ relative to the LSR does not seem to alter the shape of the CSE. This is reasonable, since it is most likely protected from immediate interaction with interstellar matter by the surrounding older shell at 155\arcsec. From what can be seen in the weak signal, it is highly symmetrically shaped. However, most of \object{RT~Cap's} space motion is in radial direction ($-20.8$\,km\,s$^{-1}$, corrected for solar motion), heading towards us. In this case, deviations from spherical symmetry caused by wind-ISM interaction are hardly traceable. Since there are no molecular line measurements and thus no velocity information for any detached part, a generic value of 15\,km\,s$^{-1}$ for the shell expansion is assumed, which makes the outer structure about 14\,200 years old. If the matter is piling up at the ISM-wind frontier, hence slowing down the wind, the corresponding mass-loss event would have occurred even earlier.

Assuming a mean ISM hydrogen density of 0.6\,cm$^{-3}$, adopting the galactic model of \citet{Loup1993}, the stellar wind would have swept up a mass of $8.9\times10^{-4}\,\mathrm{M}_{\sun}$ at a distance of 0.21\,pc. Since the ISM density in the galaxy varies strongly at smaller scales, this is, however, only a highly uncertain estimate. If we take the total mass estimated for the outer shell -- $1.2\times10^{-3}\,\mathrm{M}_{\sun}$ assuming a gas-to-dust ratio of 200 -- the shell and swept-up masses turn out to be of the same order of magnitude. In this case, the stellar wind should have already been slowed down and hence the age of the responsible mass-loss event given above would only be a lower limit.

\subsection{Modelling uncertainties}
\label{errordisc}
The error in temperature at the inner border of the detached shell is only estimated from the uncertainty of the fitted shell position, which can be constrained very well. Because the temperature gradient at the respective radii is already very flat, the calculated formal error is only $\pm1$\,K, which is not a realistic value. The main error arises from the uncertainties in stellar temperature and from the applied optical dust properties. Using other optical constants for amorphous carbon than those from \citet{Rouleau1991}, for example, the values of \citet{Preibisch1993}, $T_\mathrm{eff}$ typically changes $\pm3-4$\,K. Changing the grain size also affects the obtained temperature at similar scales, where smaller particle sizes yield a lower $T_\mathrm{eff}$ of the dust. However, the fit results do not clearly favour a particular grain size distribution. The uncertainty of the stellar temperature was not included in the calculations and does not contribute to the error provided for dust temperature or the luminosity. There are two other contributors to the luminosity error. For one, the errorbars of the individual datapoints (see Tab.~\ref{tab:photdata}) are taken into account. The error is also increased by the magnitude spread of points within a single filter because several measurements are taken at different phases of the pulsational cycle. Another potential source of error, the uncertainty of the distance of the target, was not considered because it is difficult to estimate.

By scaling the error bars of the PACS intensity profiles and the spectra, one can adjust the weight that is given to the parameters of the fit. Typically, the shape parameters, mainly determined by the intensity profiles, are too loosely confined. Reducing the errorbars of the PACS measurements improved the fit to the radial intensity distribution, without the spectral fit becoming noticeably worse. The shell position $y_1$ can be well constrained, to a lesser extent its width $\delta y$ and amplitude $s_1$, which are  anti-correlated. The uncertainty in $\delta y$ directly influences our estimate of mass-loss duration and consequently the mass-loss rate. On the other hand, the dust mass estimate for the detached shell is only mildly sensitive to the shape parameters and highly affected by the grain properties.

\subsection{Formation scenarios}  
In the two cases we presented, a short period of enhanced mass-loss is the most probable primary cause of the shell formation, very likely tied to a thermal pulse \citep[see, e.g.][]{Schoier2005,Mattsson2007,Maercker2010}. \citet{Abia2002} found technetium ($^{99}$Tc) in S Sct from a
spectral synthesis of high-resolution spectra, indicating the recent occurrence of a thermal pulse, which adds support to a connection between these phenomena. Our derived formation time-scales for such a scenario are at the upper limit of what is typically expected for the duration of such events \citep{Vassiliadis1993}. However, as discussed in Sect.~\ref{disc:ssct}, the actual times might be artificially spread because of (a) a velocity dispersion of dust grains inside the shell and (b) simplified model assumptions, meaning   that azimuthal variations are averaged into a 1D profile.

Additionally, other effects can be expected to influence the evolution of the structure. For example, \citet{Schoier2005} discussed indications for the fast massive ejection running into an older, slower wind
for known detached-shell objects. Correlations such as growing shell mass with increasing radius or a slowing down of expansion velocity indicate that such an interaction is taking place. Modelling attempts \citep[e.g.][]{Steffen2000,Mattsson2007} also favour a density-enhancing wind-wind interplay. For S~Sct, the co-spatiality of the gas and dust supports this scenario. However, a minor drift cannot be excluded because of the low resolution of the molecular line data.

The cause of the observed asymmetries remains unclear for both objects. While interaction with the ISM can be excluded because
of the enclosing second shells, density variations in the old ejecta into which the younger wind is expanding might cause the inhomogeneity. On the other hand, the mass-loss itself might be anisotropic from the beginning, as was suggested to occur especially during episodes of high MLR \citep[e.g.][]{Woitke2006}. It is especially puzzling that the regions of enhanced brightness around RT~Cap are almost complementary in the blue and red band (cf.~Figure~\ref{fig:rtcap70}). This might be indicative of temperature variations across the shell. For S~Sct there may exist a velocity shear between the young and the old wind, which might push material in the direction opposite to the space motion.

The molecular counterpart to the dust shell as far out as some $10^{17}$\,cm from S~Sct cannot be explained by the high mass-loss rate alone. As suggested  for instance by \citet{Bergman1993}, a clumped medium enhances the resistance of CO molecules against photodissociation by interstellar UV photons through self-shielding. High-resolution interferometric observations showed that the circumstellar medium can be clumpy to a high degree,  for example,
by \citet{Lindqvist1999} for \object{U~Cam} and \citet{Olofsson2000} for \object{TT~Cyg}, or more recently by \citet{Vlemmings2013} for \object{R~Scl} with ALMA. The responsible instabilities in the outflow were demonstrated by gasdynamical modelling in \citet{Myasnikov2000}. While for \object{S~Sct} there are indications for inhomogeneously distributed material in the patchy structure of the CO observations \citep{Olofsson1992}, there is no morphological evidence observed in the spatially higher resolved FIR data. There clearly are small-scale structures in the shell, but for a definite identification of dust clumps the spatial resolution of the Herschel data is still not sufficient. Although of very similar physical size as S~Sct (0.13\,pc), no CO counterpart was so far detected for the detached shell around RT~Cap. The considerably lower dust mass of the shell of RT~Cap is a possible explanation if we assume similar dust-to-gas ratios for both targets.

Two mechanisms are very probably responsible for the formation
of the extended structures that are located at distances of approximately 130$\arcsec$ (0.25\,pc) and 155$\arcsec$ (0.21\,pc) from S~Sct and RT~Cap. Interaction of an old stellar wind with the surrounding ISM, as discussed by \citet{Wareing2006}, seems very probable particularly for S Sct, given the shape and kinematic information. However, while the derived space motion of \object{S~Sct} is well above the sonic speed in the ISM ($\sim1-10$\,km\,s$^{-1}$), the objects found in clearly noticeable interaction with the interstellar environment usually move at much higher velocities, for instance, \object{$o$~Ceti} \citep{Mayer2011} and the fermata-class objects presented in \citet{Cox2012}. At 17.7\,km\,s$^{-1}$ \object{S~Sct} also travels more slowly through the ISM than the average galactic AGB star \citep[$\sim30$\,km\,s$^{-1}$,][]{Feast2000}. Therefore
the observed density enhancement is most
likely additionally influenced by an increased mass-loss rate and hence a density modulation in the stellar wind itself. This is also particularly probable for the outer envelope of RT~Cap.

\section{Conclusion}
We have investigated the circumstellar dust structures around the carbon stars S~Sct and RT~Cap. We detected a prominent detached shell that surrounds both objects, plus so far unknown older remains of either a continuous wind or even a mass eruption. While S~Sct was known to have a detached shell from CO molecular line observations, the detection of the circumstellar structures around RT~Cap is a first detection. The new observations that
reveal the two-shell morphology support the  scenario that the detached shells are a consequence of wind-wind interaction. We derived dust masses based on both new Herschel FIR observations and archive data. Moreover, thanks to the high spatial resolution PACS maps, we estimated the corresponding formation time-scales and mass-loss rates, or at least provided an upper limit. Concerning the connection of the observed detached structures to thermal pulses, the time intervals we estimated between the young and old shells seem compatible with the typical duration of an interpulse period.

The uncertain optical dust properties are a non-negligible error source in the calculation. Here a more detailed study using the various optical constants datasets for carbonaceous dust grains and different grain size distributions is desirable.

In the MESS sample we found additional stars surrounded by (newly detected) detached dust shells, which will be presented in an upcoming paper.

\begin{acknowledgements}
MM, FK and AM acknowledge funding by the Austrian Science Fund FWF under project number P23586 and FFG grant FA538019. MG, JB, LD, and BvdB acknowledge support by an ESA-Prodex grant.
\end{acknowledgements}

\bibliographystyle{aa}
\bibliography{pap2}

\begin{thebibliography}{76}
\expandafter\ifx\csname natexlab\endcsname\relax\def\natexlab#1{#1}\fi

\bibitem[{Abia {et~al.}(2002)Abia, Dom\'{i}nguez, Gallino, Busso, Masera,
  Straniero, de~Laverny, Plez, \& Isern}]{Abia2002}
Abia, C., Dom\'{i}nguez, I., Gallino, R., {et~al.} 2002, \apj, 579, 817

\bibitem[{{Arenou} {et~al.}(1992){Arenou}, {Grenon}, \& {Gomez}}]{Arenou1992}
{Arenou}, F., {Grenon}, M., \& {Gomez}, A. 1992, \aap, 258, 104

\bibitem[{{Aringer} {et~al.}(2009){Aringer}, {Girardi}, {Nowotny}, {Marigo}, \&
  {Lederer}}]{Aringer2009}
{Aringer}, B., {Girardi}, L., {Nowotny}, W., {Marigo}, P., \& {Lederer}, M.~T.
  2009, \aap, 503, 913

\bibitem[{{Beichman} {et~al.}(1988){Beichman}, {Neugebauer}, {Habing}, {Clegg},
  \& {Chester}}]{Beichman1988}
{Beichman}, C.~A., {Neugebauer}, G., {Habing}, H.~J., {Clegg}, P.~E., \&
  {Chester}, T.~J., eds. 1988, {Infrared astronomical satellite (IRAS) catalogs
  and atlases. Volume 1: Explanatory supplement}, Vol.~1

\bibitem[{Bergman {et~al.}(1993)Bergman, Carlstrom, \& Olofsson}]{Bergman1993}
Bergman, P., Carlstrom, U., \& Olofsson, H. 1993, \aap, 268, 685

\bibitem[{{Cox} {et~al.}(2012){Cox}, {Kerschbaum}, {van Marle}, {Decin},
  {Ladjal}, {Mayer}, {Groenewegen}, {van Eck}, {Royer}, {Ottensamer}, {Ueta},
  {Jorissen}, {Mecina}, {Meliani}, {Luntzer}, {Blommaert}, {Posch},
  {Vandenbussche}, \& {Waelkens}}]{Cox2012}
{Cox}, N.~L.~J., {Kerschbaum}, F., {van Marle}, A.-J., {et~al.} 2012, \aap,
  537, A35

\bibitem[{{Cutri} {et~al.}(2003){Cutri}, {Skrutskie}, {van Dyk}, {Beichman},
  {Carpenter}, {Chester}, {Cambresy}, {Evans}, {Fowler}, {Gizis}, {Howard},
  {Huchra}, {Jarrett}, {Kopan}, {Kirkpatrick}, {Light}, {Marsh}, {McCallon},
  {Schneider}, {Stiening}, {Sykes}, {Weinberg}, {Wheaton}, {Wheelock}, \&
  {Zacarias}}]{Cutri2003}
{Cutri}, R.~M., {Skrutskie}, M.~F., {van Dyk}, S., {et~al.} 2003, VizieR Online
  Data Catalog, 2246, 0

\bibitem[{{Cutri} {et~al.}(2012){Cutri}, {Wright}, {Conrow}, {Bauer},
  {Benford}, {Brandenburg}, {Dailey}, {Eisenhardt}, {Evans}, {Fajardo-Acosta},
  {Fowler}, {Gelino}, {Grillmair}, {Harbut}, {Hoffman}, {Jarrett},
  {Kirkpatrick}, {Leisawitz}, {Liu}, {Mainzer}, {Marsh}, {Masci}, {McCallon},
  {Padgett}, {Ressler}, {Royer}, {Skrutskie}, {Stanford}, {Wyatt}, {Tholen},
  {Tsai}, {Wachter}, {Wheelock}, {Yan}, {Alles}, {Beck}, {Grav}, {Masiero},
  {McCollum}, {McGehee}, {Papin}, \& {Wittman}}]{Cutri2012}
{Cutri}, R.~M., {Wright}, E.~L., {Conrow}, T., {et~al.} 2012, {Explanatory
  Supplement to the WISE All-Sky Data Release Products}, Tech. rep.

\bibitem[{Drimmel {et~al.}(2003)Drimmel, Cabrera-Lavers, \&
  L\'{o}pez-Corredoira}]{Drimmel2003}
Drimmel, R., Cabrera-Lavers, A., \& L\'{o}pez-Corredoira, M. 2003, \aap, 409,
  205

\bibitem[{Eriksson \& Stenholm(1993)}]{Eriksson1993}
Eriksson, K. \& Stenholm, L. 1993, \aap, 271, 508

\bibitem[{{Feast} \& {Whitelock}(2000)}]{Feast2000}
{Feast}, M.~W. \& {Whitelock}, P.~A. 2000, \mnras, 317, 460

\bibitem[{Geise {et~al.}(2010)Geise, Ueta, Speck, Izumiura, \&
  Stencel}]{Geise2010}
Geise, K.~M., Ueta, T., Speck, A.~K., Izumiura, H., \& Stencel, R.~E. 2010, in
  American Astronomical Society Meeting Abstracts {\textbackslash}\#215,
  Vol.~42, 364

\bibitem[{{Groenewegen}(2008)}]{Groenewegen2008}
{Groenewegen}, M.~A.~T. 2008, \aap, 488, 935

\bibitem[{{Groenewegen}(2012)}]{Groenewegen2012}
{Groenewegen}, M.~A.~T. 2012, \aap, 543, A36

\bibitem[{Groenewegen \& de~Jong(1994)}]{Groenewegen1994}
Groenewegen, M. A.~T. \& de~Jong, T. 1994, \aap, 282, 115

\bibitem[{Groenewegen {et~al.}(2011)Groenewegen, Waelkens, Barlow, Kerschbaum,
  Garcia-Lario, Cernicharo, Blommaert, Bouwman, Cohen, Cox, Decin, Exter, Gear,
  Gomez, Hargrave, Henning, Hutsem\'{e}kers, Ivison, Jorissen, Krause, Ladjal,
  Leeks, Lim, Matsuura, Naz\'{e}, Olofsson, Ottensamer, Polehampton, Posch,
  Rauw, Royer, Sibthorpe, Swinyard, Ueta, Vamvatira-Nakou, Vandenbussche,
  van~de Steene, van Eck, van Hoof, van Winckel, Verdugo, \&
  Wesson}]{Groenewegen2011}
Groenewegen, M. A.~T., Waelkens, C., Barlow, M.~J., {et~al.} 2011, \aap, 526,
  162

\bibitem[{Habing(1996)}]{Habing1996}
Habing, H.~J. 1996, \aapr, 7, 97

\bibitem[{Ishihara {et~al.}(2010)Ishihara, Onaka, Kataza, Salama, Alfageme,
  Cassatella, Cox, Garc\'{i}a-Lario, Stephenson, Cohen, Fujishiro, Fujiwara,
  Hasegawa, Ita, Kim, Matsuhara, Murakami, M\"{u}ller, Nakagawa, Ohyama, Oyabu,
  Pyo, Sakon, Shibai, Takita, Tanab\'{e}, Uemizu, Ueno, Usui, Wada, Watarai,
  Yamamura, \& Yamauchi}]{Ishihara2010}
Ishihara, D., Onaka, T., Kataza, H., {et~al.} 2010, \aap, 514, 1

\bibitem[{Ivezic {et~al.}(1999)Ivezic, Nenkova, \& Elitzur}]{Ivezic1999}
Ivezic, Z., Nenkova, M., \& Elitzur, M. 1999, User Manual for {DUSTY}

\bibitem[{Izumiura {et~al.}(1996)Izumiura, Hashimoto, Kawara, Yamamura, \&
  Waters}]{Izumiura1996}
Izumiura, H., Hashimoto, O., Kawara, K., Yamamura, I., \& Waters, L. B. F.~M.
  1996, \aap, 315, L221

\bibitem[{Izumiura {et~al.}(2009)Izumiura, Ueta, Yamamura, Nakada, Matsunaga,
  Ita, Matsuura, Fukushi, Mito, \& Tanabe}]{Izumiura2009}
Izumiura, H., Ueta, T., Yamamura, I., {et~al.} 2009, in {AKARI}, a Light to
  Illuminate the Misty Universe, Vol. 418, 127

\bibitem[{{Johnson} \& {Soderblom}(1987)}]{Johnson1987}
{Johnson}, D.~R.~H. \& {Soderblom}, D.~R. 1987, \aj, 93, 864

\bibitem[{{Kerschbaum} \& {Hron}(1994)}]{Kerschbaum1994}
{Kerschbaum}, F. \& {Hron}, J. 1994, \aaps, 106, 397

\bibitem[{Kerschbaum {et~al.}(2010)Kerschbaum, Ladjal, Ottensamer, Groenewegen,
  Mecina, Blommaert, Baumann, Decin, Vandenbussche, Waelkens, Posch, Huygen,
  De~Meester, Regibo, Royer, Exter, \& Jean}]{Kerschbaum2010}
Kerschbaum, F., Ladjal, D., Ottensamer, R., {et~al.} 2010, \aap, 518, L140

\bibitem[{{Kharchenko} \& {Roeser}(2009)}]{Kharchenko2009}
{Kharchenko}, N.~V. \& {Roeser}, S. 2009, VizieR Online Data Catalog, 1280, 0

\bibitem[{{Lambert} {et~al.}(1986){Lambert}, {Gustafsson}, {Eriksson}, \&
  {Hinkle}}]{Lambert1986}
{Lambert}, D.~L., {Gustafsson}, B., {Eriksson}, K., \& {Hinkle}, K.~H. 1986,
  \apjs, 62, 373

\bibitem[{Lindqvist {et~al.}(1996)Lindqvist, Lucas, Olofsson, Omont, Eriksson,
  \& Gustafsson}]{Lindqvist1996}
Lindqvist, M., Lucas, R., Olofsson, H., {et~al.} 1996, \aap, 305, L57

\bibitem[{Lindqvist {et~al.}(1999)Lindqvist, Olofsson, Lucas, Sch\"{o}ier,
  Neri, Bujarrabal, \& Kahane}]{Lindqvist1999}
Lindqvist, M., Olofsson, H., Lucas, R., {et~al.} 1999, \aap, 351, L1

\bibitem[{{Loup} {et~al.}(1993){Loup}, {Forveille}, {Omont}, \&
  {Paul}}]{Loup1993}
{Loup}, C., {Forveille}, T., {Omont}, A., \& {Paul}, J.~F. 1993, \aaps, 99, 291

\bibitem[{{Maercker} {et~al.}(2012){Maercker}, {Mohamed}, {Vlemmings},
  {Ramstedt}, {Groenewegen}, {Humphreys}, {Kerschbaum}, {Lindqvist},
  {Olofsson}, {Paladini}, {Wittkowski}, {de Gregorio-Monsalvo}, \&
  {Nyman}}]{Maercker2012}
{Maercker}, M., {Mohamed}, S., {Vlemmings}, W.~H.~T., {et~al.} 2012, \nat, 490,
  232

\bibitem[{Maercker {et~al.}(2010)Maercker, Olofsson, Eriksson, Gustafsson, \&
  Sch\"{o}ier}]{Maercker2010}
Maercker, M., Olofsson, H., Eriksson, K., Gustafsson, B., \& Sch\"{o}ier, F.~L.
  2010, \aap, 511, 37

\bibitem[{{Marshall} {et~al.}(2006){Marshall}, {Robin}, {Reyl{\'e}},
  {Schultheis}, \& {Picaud}}]{Marshall2006}
{Marshall}, D.~J., {Robin}, A.~C., {Reyl{\'e}}, C., {Schultheis}, M., \&
  {Picaud}, S. 2006, \aap, 453, 635

\bibitem[{{Mattsson} {et~al.}(2007){Mattsson}, {H{\"o}fner}, \&
  {Herwig}}]{Mattsson2007}
{Mattsson}, L., {H{\"o}fner}, S., \& {Herwig}, F. 2007, \aap, 470, 339

\bibitem[{{Mattsson} {et~al.}(2010){Mattsson}, {Wahlin}, \&
  {H{\"o}fner}}]{Mattsson2010}
{Mattsson}, L., {Wahlin}, R., \& {H{\"o}fner}, S. 2010, \aap, 509, A14,
  provided by the SAO/NASA Astrophysics Data System

\bibitem[{{Mayer} {et~al.}(2011){Mayer}, {Jorissen}, {Kerschbaum}, {Mohamed},
  {van Eck}, {Ottensamer}, {Blommaert}, {Decin}, {Groenewegen}, {Posch},
  {Vandenbussche}, \& {Waelkens}}]{Mayer2011}
{Mayer}, A., {Jorissen}, A., {Kerschbaum}, F., {et~al.} 2011, \aap, 531, L4

\bibitem[{Min {et~al.}(2003)Min, Hovenier, \& de~Koter}]{Min2003}
Min, M., Hovenier, J.~W., \& de~Koter, A. 2003, \aap, 404, 35

\bibitem[{{Min} {et~al.}(2005){Min}, {Hovenier}, \& {de Koter}}]{Min2005}
{Min}, M., {Hovenier}, J.~W., \& {de Koter}, A. 2005, \aap, 432, 909, provided
  by the SAO/NASA Astrophysics Data System

\bibitem[{{Myasnikov} {et~al.}(2000){Myasnikov}, {Belov}, {Gustafsson}, \&
  {Eriksson}}]{Myasnikov2000}
{Myasnikov}, A.~V., {Belov}, N.~A., {Gustafsson}, B., \& {Eriksson}, K. 2000,
  \apss, 274, 231, provided by the SAO/NASA Astrophysics Data System

\bibitem[{Noguchi {et~al.}(1981)Noguchi, Kawara, Kobayashi, Okuda, Sato, \&
  Oishi}]{Noguchi1981}
Noguchi, K., Kawara, K., Kobayashi, Y., {et~al.} 1981, \pasj, 33, 373

\bibitem[{Olofsson {et~al.}(1996)Olofsson, Bergman, Eriksson, \&
  Gustafsson}]{Olofsson1996}
Olofsson, H., Bergman, P., Eriksson, K., \& Gustafsson, B. 1996, \aap, 311, 587

\bibitem[{Olofsson {et~al.}(2000)Olofsson, Bergman, Lucas, Eriksson,
  Gustafsson, \& Bieging}]{Olofsson2000}
Olofsson, H., Bergman, P., Lucas, R., {et~al.} 2000, \aap, 353, 583

\bibitem[{Olofsson {et~al.}(1992)Olofsson, Carlstrom, Eriksson, \&
  Gustafsson}]{Olofsson1992}
Olofsson, H., Carlstrom, U., Eriksson, K., \& Gustafsson, B. 1992, \aap, 253,
  L17

\bibitem[{Olofsson {et~al.}(1990)Olofsson, Carlstrom, Eriksson, Gustafsson, \&
  Willson}]{Olofsson1990}
Olofsson, H., Carlstrom, U., Eriksson, K., Gustafsson, B., \& Willson, L.~A.
  1990, \aap, 230, L13

\bibitem[{{Olofsson} {et~al.}(1988){Olofsson}, {Eriksson}, \&
  {Gustafsson}}]{Olofsson1988}
{Olofsson}, H., {Eriksson}, K., \& {Gustafsson}, B. 1988, \aap, 196, L1

\bibitem[{Olofsson {et~al.}(1993)Olofsson, Eriksson, Gustafsson, \&
  Carlstroem}]{Olofsson1993}
Olofsson, H., Eriksson, K., Gustafsson, B., \& Carlstroem, U. 1993, \apjs, 87,
  305

\bibitem[{{Ottensamer} {et~al.}(2011){Ottensamer}, {Luntzer}, {Mecina},
  {Kerschbaum}, {Blommaert}, {Decin}, {Groenewegen}, {Posch}, {Vandenbussche},
  \& {Waelkens}}]{Ottensamer2011}
{Ottensamer}, R., {Luntzer}, A., {Mecina}, M., {et~al.} 2011, in Astronomical
  Society of the Pacific Conference Series, Vol. 445, Why Galaxies Care about
  AGB Stars II: Shining Examples and Common Inhabitants, ed. F.~{Kerschbaum},
  T.~{Lebzelter}, \& R.~F. {Wing}, 625

\bibitem[{Parenago(1940)}]{Parenago1940}
Parenago, P.~P. 1940, Astron. Zh., 17, 3

\bibitem[{Pilbratt {et~al.}(2010)Pilbratt, Riedinger, Passvogel, Crone, Doyle,
  Gageur, Heras, Jewell, Metcalfe, Ott, \& Schmidt}]{Pilbratt2010}
Pilbratt, G.~L., Riedinger, J.~R., Passvogel, T., {et~al.} 2010, \aap, 518, L1

\bibitem[{{Pitman} {et~al.}(2008){Pitman}, {Hofmeister}, {Corman}, \&
  {Speck}}]{Pitman2008}
{Pitman}, K.~M., {Hofmeister}, A.~M., {Corman}, A.~B., \& {Speck}, A.~K. 2008,
  \aap, 483, 661

\bibitem[{Poglitsch {et~al.}(2010)Poglitsch, Waelkens, Geis, Feuchtgruber,
  Vandenbussche, Rodriguez, Krause, Renotte, van Hoof, Saraceno, Cepa,
  Kerschbaum, Agn\`{e}se, Ali, Altieri, Andreani, Augueres, Balog, Barl, Bauer,
  Belbachir, Benedettini, Billot, Boulade, Bischof, Blommaert, Callut, Cara,
  Cerulli, Cesarsky, Contursi, Creten, De~Meester, Doublier, Doumayrou, Duband,
  Exter, Genzel, Gillis, Gr\"{o}zinger, Henning, Herreros, Huygen, Inguscio,
  Jakob, Jamar, Jean, de~Jong, Katterloher, Kiss, Klaas, Lemke, Lutz, Madden,
  Marquet, Martignac, Mazy, Merken, Montfort, Morbidelli, M\"{u}ller, Nielbock,
  Okumura, Orfei, Ottensamer, Pezzuto, Popesso, Putzeys, Regibo, Reveret,
  Royer, Sauvage, Schreiber, Stegmaier, Schmitt, Schubert, Sturm, Thiel,
  Tofani, Vavrek, Wetzstein, Wieprecht, \& Wiezorrek}]{Poglitsch2010}
Poglitsch, A., Waelkens, C., Geis, N., {et~al.} 2010, \aap, 518, L2

\bibitem[{{Preibisch} {et~al.}(1993){Preibisch}, {Ossenkopf}, {Yorke}, \&
  {Henning}}]{Preibisch1993}
{Preibisch}, T., {Ossenkopf}, V., {Yorke}, H.~W., \& {Henning}, T. 1993, \aap,
  279, 577

\bibitem[{{Ramstedt} {et~al.}(2011){Ramstedt}, {Maercker}, {Olofsson},
  {Olofsson}, \& {Sch{\"o}ier}}]{Ramstedt2011}
{Ramstedt}, S., {Maercker}, M., {Olofsson}, G., {Olofsson}, H., \&
  {Sch{\"o}ier}, F.~L. 2011, \aap, 531, A148

\bibitem[{{Rouleau} \& {Martin}(1991)}]{Rouleau1991}
{Rouleau}, F. \& {Martin}, P.~G. 1991, \apj, 377, 526

\bibitem[{{Roussel}(2013)}]{Roussel2013}
{Roussel}, H. 2013, \pasp, 125, 1126, provided by the SAO/NASA Astrophysics
  Data System

\bibitem[{Samus {et~al.}(2012)Samus, Durlevich, Kazarovets, Kireeva,
  Pastukhova, \& Zharova}]{Samus2012}
Samus, N., Durlevich, O., Kazarovets, E.~V., {et~al.} 2012, GCVS database

\bibitem[{Sch\"{o}ier {et~al.}(2005)Sch\"{o}ier, Lindqvist, \&
  Olofsson}]{Schoier2005}
Sch\"{o}ier, F.~L., Lindqvist, M., \& Olofsson, H. 2005, \aap, 436, 633

\bibitem[{{Sch{\"o}ier} \& {Olofsson}(2001)}]{Schoier2001}
{Sch{\"o}ier}, F.~L. \& {Olofsson}, H. 2001, \aap, 368, 969

\bibitem[{Schr\"{o}der \& Sedlmayr(2001)}]{Schroder2001}
Schr\"{o}der, K.-P. \& Sedlmayr, E. 2001, \aap, 366, 913

\bibitem[{Schwarz(1978)}]{Schwarz1978}
Schwarz, G. 1978, The Annals of Statistics, 6, pp. 461

\bibitem[{{Sloan} {et~al.}(2003){Sloan}, {Kraemer}, {Price}, \&
  {Shipman}}]{Sloan2003}
{Sloan}, G.~C., {Kraemer}, K.~E., {Price}, S.~D., \& {Shipman}, R.~F. 2003,
  \apjs, 147, 379

\bibitem[{{Steffen} \& {Sch{\"o}nberner}(2000)}]{Steffen2000}
{Steffen}, M. \& {Sch{\"o}nberner}, D. 2000, \aap, 357, 180

\bibitem[{{Ueta} {et~al.}(2008){Ueta}, {Izumiura}, {Yamamura}, {Nakada},
  {Matsuura}, {Ita}, {Tanab{\'e}}, {Fukushi}, {Matsunaga}, \&
  {Mito}}]{Ueta2008}
{Ueta}, T., {Izumiura}, H., {Yamamura}, I., {et~al.} 2008, \pasj, 60, 407

\bibitem[{{Ueta} {et~al.}(2006){Ueta}, {Speck}, {Stencel}, {Herwig}, {Gehrz},
  {Szczerba}, {Izumiura}, {Zijlstra}, {Latter}, {Matsuura}, {Meixner},
  {Steffen}, \& {Elitzur}}]{Ueta2006}
{Ueta}, T., {Speck}, A.~K., {Stencel}, R.~E., {et~al.} 2006, \apjl, 648, L39

\bibitem[{van~der Veen \& Habing(1988)}]{Veen1988}
van~der Veen, W. E. C.~J. \& Habing, H.~J. 1988, \aap, 194, 125

\bibitem[{van Leeuwen(2007)}]{Leeuwen2007}
van Leeuwen, F. 2007, \aap, 474, 653

\bibitem[{Vassiliadis \& Wood(1993)}]{Vassiliadis1993}
Vassiliadis, E. \& Wood, P.~R. 1993, \apj, 413, 641

\bibitem[{{Vlemmings} {et~al.}(2013){Vlemmings}, {Maercker}, {Lindqvist},
  {Mohamed}, {Olofsson}, {Ramstedt}, {Brunner}, {Groenewegen}, {Kerschbaum}, \&
  {Wittkowski}}]{Vlemmings2013}
{Vlemmings}, W.~H.~T., {Maercker}, M., {Lindqvist}, M., {et~al.} 2013, \aap,
  556, L1

\bibitem[{{Volk} \& {Cohen}(1989)}]{Volk1989}
{Volk}, K. \& {Cohen}, M. 1989, \aj, 98, 931, provided by the SAO/NASA
  Astrophysics Data System

\bibitem[{Walker(1980)}]{Walker1980}
Walker, A.~R. 1980, \mnras, 190, 543

\bibitem[{{Wareing} {et~al.}(2006){Wareing}, {Zijlstra}, {Speck}, {O'Brien},
  {Ueta}, {Elitzur}, {Gehrz}, {Herwig}, {Izumiura}, {Matsuura}, {Meixner},
  {Stencel}, \& {Szczerba}}]{Wareing2006}
{Wareing}, C.~J., {Zijlstra}, A.~A., {Speck}, A.~K., {et~al.} 2006, \mnras,
  372, L63

\bibitem[{Whitelock {et~al.}(2006)Whitelock, Feast, Marang, \&
  Groenewegen}]{Whitelock2006}
Whitelock, P.~A., Feast, M.~W., Marang, F., \& Groenewegen, M. A.~T. 2006,
  \mnras, 369, 751

\bibitem[{{Woitke}(2006)}]{Woitke2006}
{Woitke}, P. 2006, \aap, 452, 537

\bibitem[{Wong {et~al.}(2004)Wong, Sch\"{o}ier, Lindqvist, \&
  Olofsson}]{Wong2004}
Wong, T., Sch\"{o}ier, F.~L., Lindqvist, M., \& Olofsson, H. 2004, \aap, 413,
  241

\bibitem[{Yamamura {et~al.}(2010)Yamamura, Makiuti, Ikeda, Fukuda, Oyabu, T.,
  \& {G.J.}}]{Yamamura2010}
Yamamura, I., Makiuti, S., Ikeda, N., {et~al.} 2010, {ISAS/JAXA} {(viZier}
  catalog {II/298)}

\bibitem[{Yamamura {et~al.}(1993)Yamamura, Onaka, Kamijo, Izumiura, \&
  Deguchi}]{Yamamura1993}
Yamamura, I., Onaka, T., Kamijo, F., Izumiura, H., \& Deguchi, S. 1993, \pasj,
  45, 573

\bibitem[{{Young} {et~al.}(1993){Young}, {Phillips}, \& {Knapp}}]{Young1993}
{Young}, K., {Phillips}, T.~G., \& {Knapp}, G.~R. 1993, \apjs, 86, 517

\end{thebibliography}

\newpage
\emph{}
\newpage
\appendix
\section{Observation data}
\label{apdx1}
\begin{figure}
\centering
\includegraphics[width=0.9\hsize]{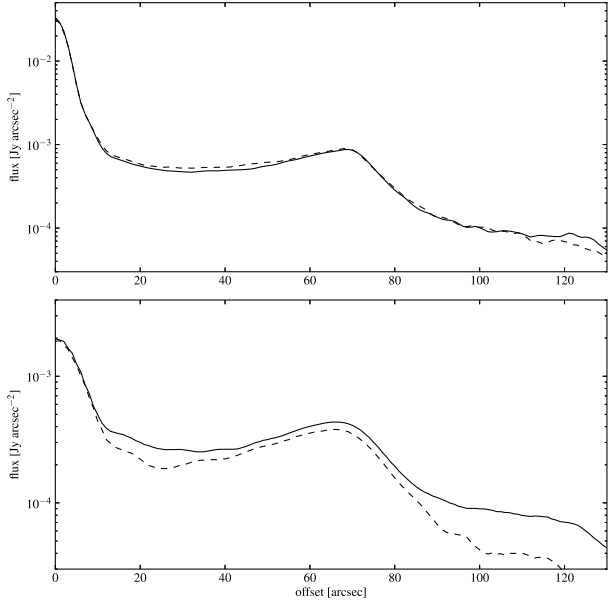}
\caption{Azimuthally averaged radial profiles of \object{S~Sct}. \emph{Top:} the first (dashed) and the re-observation (solid - adopted for the modelling) in the blue filter. \emph{Bottom:} corresponding profiles in the red band.}
\label{fig:ssct_comp}
\end{figure}
\begin{figure}
\centering
\includegraphics[width=0.9\hsize,trim=0 0 0 5,clip]{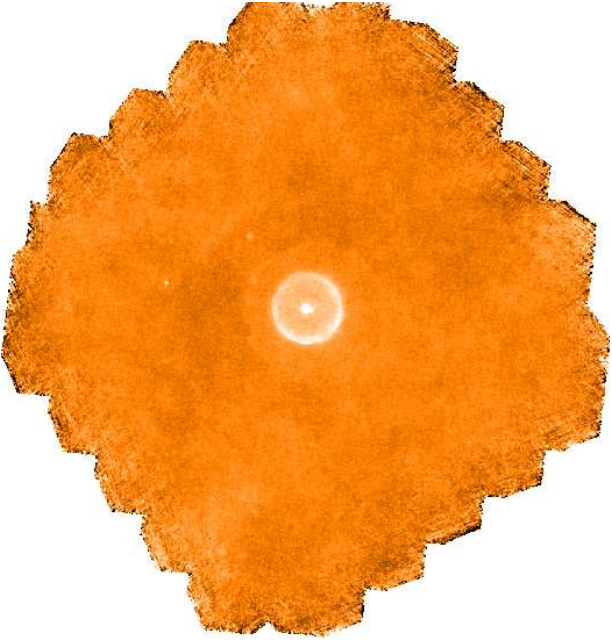}
\caption{OBSID 1342219068 and 1342219069 observation of S~Sct in the blue band.}
\label{fig:ssct_old}
\end{figure}
\begin{figure}
\centering
\includegraphics[width=0.9\hsize,trim=0 0 0 5,clip]{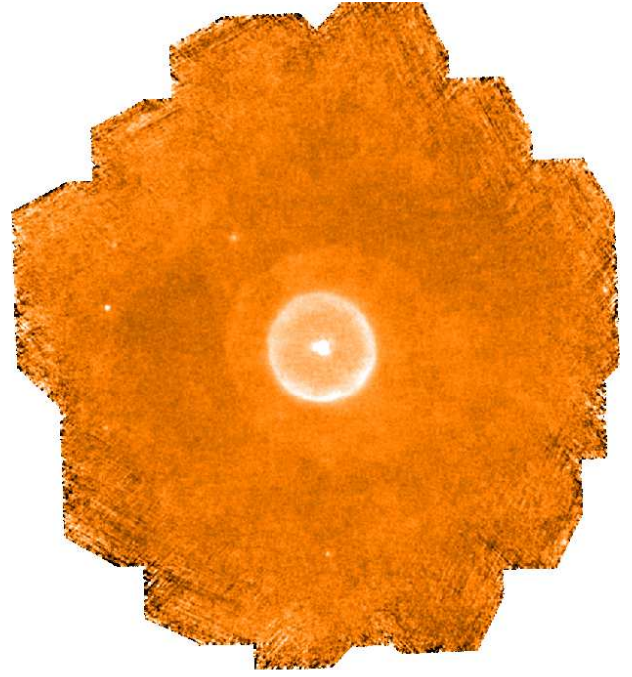}
\caption{OBSID 1342229077 and 1342229078 observation of S~Sct in the blue band. The field of view and scale are different from
those of OD 705.}
\label{fig:ssct_new}
\end{figure}
In principle, a combination of all available Herschel PACS data (i.e. both observing dates) is desirable to recover the faintest structures as well as possible. There are, however, significant differences in the pointing between the first and the re-observation, which result in a smeared image if the datasets are combined uncorrected. A modification of the astrometric information is possible by hand, but the results are not fully satisfactory. In the case of \object{RT~Cap} the use of only the better (i.e. higher SNR) dataset is justified, since the addition of the much shorter observation \citep[see, e.g.,][]{Cox2012} does not improve the final map noticeably. For \object{S Sct} the case is different, since both datasets have similar depths. Although the individual observations by themselves have very good signal-to-noise ratios
(S/N), a combination of all available data does not reveal additional morphological information and does not give any benefits regarding the azimuthally averaged profiles used for modelling. Therefore we eventually used here only a single observation, OBSID 1342229077 and 1342229078, because of its better coverage resulting from the smaller scanned area at similar observing duration. A juxtaposition of the S~Sct observations is given in the following.

Observations from both ODs were reduced using HIPE and mapped with Scanamorphos version 13, which means that the data processing and calibration can be assumed to be consistent (the full maps are shown in Figs.~\ref{fig:ssct_old}\, and\,\ref{fig:ssct_new}). However, the numbers in Table~\ref{tab:ssct_photcomp} show some differences in the photometry. In the blue band the discrepancy is about 6\%, in the red band it is 9\%. While at $70\,\mu$m the observation taken on OD 705 suggests a higher flux than on OD 860, the situation  is just the reverse for the $160\,\mu$m measurement. Nonetheless, the relative errors between the two bands more or less comply with the expected photometric accuracies. The largest uncertainty of the flux calibration is the background estimation performed by Scanamorphos. The differences shown in Fig.~\ref{fig:ssct_comp}, especially visible in the red, are partly caused by the varying background subtraction in the two cases. Considering the rather poor model fits to the brightness
distribution of \object{S~Sct} , the impact of the slightly different observed profiles on the modelling outcome is negligible.

\begin{table}
\centering
\caption{Annular aperture photometry of the different ODs. Aperture radius: 80$\arcsec$, background: 90-115$\arcsec$.}
\label{tab:ssct_photcomp}
\begin{tabular}{lcc}
\hline
\hline
 & $F_{70}$ & $F_{160}$ \\
\hline
 OD 705 & 12.3 Jy & 4.7 Jy \\
 OD 860 & 11.6 Jy & 5.1 Jy \\
\hline
\end{tabular}
\end{table}

\Online

\onecolumn
\section{Photometric data}
\label{photdata}
\begin{table*}[!h]
\caption{Photometric data for S~Sct and RT~Cap. All bands are given in magnitudes, except for the two PACS filters, which are in mJy. AKARI and 2MASS errors are taken from the catalogues, UBVRIJHK errors are assumed.}
\label{tab:photdata}
\centering
\begin{tabular}{lrr|lrr}
\multicolumn{3}{c}{S Sct} & \multicolumn{3}{c}{RT Cap}\\
filter & magnitude & error & filter & magnitude & error\\
\hline
pacsb & 11600.000 & 1160.000 & pacsb & 3980.000 & 398.000 \\
pacsr & 5100.000 & 765.000 & pacsr & 1640.000 & 246.000 \\
2massJ & 2.303 & 0.314 & AkaS9W & -0.434 & 0.026 \\
2massH & 1.140 & 0.262 & AkL18W & -0.762 & 0.016 \\
2massK & 0.627 & 0.288 & AkaN60 & -1.379 & 0.206 \\
BesB & 10.23 & 0.050 & AkarWS & -1.595 & 0.115 \\
BesV & 6.93 & 0.050 & iras12 & -0.625 & 0.100 \\
BesR & 5.52 & 0.050 & iras25 & -0.847 & 0.100 \\
BesI & 4.14 & 0.050 & iras60 & -1.121 & 0.100 \\
BesB & 9.94 & 0.050 & ira100 & -2.214 & 0.100 \\
BesV & 6.82 & 0.050 & 2massJ & 2.084 & 0.244 \\
BesB & 9.90 & 0.050 & 2massH & 0.923 & 0.276 \\
BesV & 6.80 & 0.050 & 2massK & 0.305 & 0.316 \\
saaoJ & 2.43 & 0.030 & esoJ & 2.370 & 0.100 \\
saaoH & 1.13 & 0.030 & esoH & 1.330 & 0.100 \\
saaoK & 0.57 & 0.030 & esoK & 0.600 & 0.100 \\
saaoL & 0.05 & 0.050 & esoL & 0.050 & 0.100 \\
saaoJ & 2.08 & 0.040 & esoJ & 2.410 & 0.050 \\
saaoH & 1.19 & 0.040 & esoH & 1.220 & 0.050 \\
saaoK & 0.55 & 0.040 & esoK & 0.560 & 0.050 \\
saaoL & 0.09 & 0.080 & esoL & -0.160 & 0.050 \\
esoJ & 1.990 & 0.050 & esoM & 0.320 & 0.100 \\
esoH & 1.060 & 0.050 & BesB & 9.137 & 0.181 \\
esoK & 0.440 & 0.050 & BesV & 7.133 & 0.008 \\
esoL & 0.010 & 0.050 & saaoJ & 2.480 & 0.030 \\
iras12 & -0.505 & 0.050 & saaoH & 1.150 & 0.030 \\
iras25 & -0.650 & 0.050 & saaoK & 0.540 & 0.030 \\
AkaS9W & -0.329 & 0.019 & saaoL & -0.060 & 0.050 \\
AkL18W & -0.680 & 0.024 & & & \\
iras60 & -1.929 & 0.050 & & & \\
ira100 & -3.693 & 0.050 & & &
\end{tabular}
\end{table*}

\twocolumn
\end{document}